\documentclass[12pt,showpacs,showkeys,amsmath,amssymb]{revtex4}
\usepackage{amsmath,amsfonts,amsthm,amscd,amssymb,latexsym, textcomp}
\usepackage{bm}
\usepackage{dcolumn}
\usepackage{graphicx}
\usepackage{epstopdf}
\usepackage{color}
\usepackage{epsf}
\usepackage{epsfig}
\usepackage{graphicx, epic, eepic, color}
\usepackage{verbatim}
\usepackage{comment}

\date{\today}

\begin{document}

\title{Nonlocal Newtonian Cosmology}

\author{C. Chicone}
\email{chiconec@missouri.edu}
\affiliation{Department of Mathematics and Department of Physics and Astronomy, University of Missouri, Columbia,
Missouri 65211, USA }

\author{B. Mashhoon}
\email{mashhoonb@missouri.edu}
\affiliation{Department of Physics and Astronomy,
University of Missouri, Columbia, Missouri 65211, USA}

\begin{abstract} 

We explore some of  the cosmological implications of the recent classical nonlocal generalization of Einstein's theory of gravitation in which nonlocality is due to the gravitational memory of past events.  In the Newtonian regime of this theory, the nonlocal character of gravity simulates dark matter in spiral galaxies and clusters of galaxies. However, dark matter is considered indispensable as well for structure formation in standard models of cosmology. Can nonlocal gravity solve the problem of structure formation without  recourse to dark matter? Here we make a beginning in this direction by extending  nonlocal gravity in the Newtonian regime  to the cosmological domain. The nonlocal analog of the Zel'dovich solution is formulated  and the consequences of the resulting nonlocal Zel'dovich model are investigated in detail. 

\end{abstract}



\maketitle

\section{Introduction}

The theory of relativity is based on a fundamental assumption of \emph{locality}. That is, Lorentz transformations are first extended in a \emph{pointwise} manner to the measurements of accelerated observers and then to gravitational fields in general relativity via Einstein's local principle of equivalence~\cite{Ei, BM0}. On the other hand, Bohr and Rosenfeld have shown that the measurement of the classical electromagnetic field cannot be performed instantaneously and involves instead an averaging procedure over past spacetime events~\cite{B+R1, B+R2}. To go beyond the locality postulate and incorporate history dependence for fields, the past history of the accelerated observer and the fields must be taken into account. On this basis, a nonlocal special relativity theory has been developed in which the fields are local but satisfy integro-differential equations~\cite{BM1, BM2}. The deep connection between inertia and gravitation suggests that the latter should be history dependent as well. To implement this idea, a classical nonlocal generalization of Einstein's theory of gravitation has recently been developed. This  nonlocal gravity theory  is an extension of general relativity in which the past history of the gravitational field is taken into account via a certain spacetime average involving a kernel that, at this stage of the development of the theory, must be determined from observation. Nonlocal gravity is thus memory dependent. Memory dies out in space and time, features that must be reflected in the choice of the kernel of the theory. The gravitational field is local, but satisfies the integro-differential equations of nonlocal gravity theory~\cite{NL1, NL2, NL3, NL4, CM, NL5}. 

A significant feature of nonlocal gravity  is that the nonlocal aspect of gravity appears to simulate dark matter. That is, in this theory there is no dark matter; hence, what appears as dark matter in astrophysics and cosmology is expected to be due to the nonlocal  character of gravity. The persistent negative result of experiments that have searched for the particles of dark matter naturally leads to the possibility that what appears as dark matter in astronomy is in fact an aspect of the gravitational interaction. Along this line of thought, many theories have been proposed. Among such theories, nonlocal theories of gravitation form a prominent subclass. The inspiration for nonlocal gravitation theories often has its origin in developments in quantum field theory---see, for instance, Refs.~\cite{ST1,ST2,ST3} and the references cited therein. A proper consideration of such theories is well beyond the scope of the present paper, which is primarily based on a classical nonlocal generalization of Einstein's general relativity that has originated from a critical analysis of the fundamental assumption of locality that underlies the standard theory of relativity. 

The nonlocal character of gravity that is considered in this work cannot yet replace dark matter on all physical scales. Dark matter is currently required in astrophysics for explaining the gravitational dynamics of galaxies as well as clusters of galaxies~\cite{NL6}, gravitational lensing observations~\cite{NL5} and structure formation in cosmology. We emphasize that only some of the implications of nonlocal gravity theory have thus far been confronted with observation~\cite{NL6, ChMa}.
Nonlocal gravity is in the early stages of development at present and besides the Minkowski spacetime, no exact solution of the theory is known. The absence of any exact nontrivial solution of the theory implies that the nonlinear regime of the theory, including exact cosmological models as well as considerations involving dark energy, are beyond the scope of this work~\cite{BiMa}. However, the general linear approximation of nonlocal gravity beyond Minkowski spacetime has been investigated in detail~\cite{NL5}. In particular, the implications of the theory in the Newtonian regime have been shown to be consistent with gravitational dynamics in spiral galaxies, clusters of galaxies and the Solar System~\cite{NL6, ChMa}. On the other hand, in the standard models of cosmology, dark matter is absolutely necessary for cosmological structure formation. It is therefore of basic importance to investigate whether the nonlocal character of gravity can effectively replace dark matter in cosmological structure formation. To make a beginning in this direction is the main motivation for our work. 

The remarkable isotropy of the cosmic microwave background radiation indicates a small amplitude
($\delta \approx 10^{-5}$) for the inhomogeneities that must have existed at the epoch of decoupling ($z \approx 10^{3}$). The tremendous  growth of such inhomogeneities from the recombination era to the present time is due to the intrinsic gravitational instability of a nearly homogeneous distribution of matter.  The exact manner in which galaxies, clusters of galaxies and eventually the cosmic web have come about is not known; however, it is generally  believed that \emph{dark matter} has played a crucial role in this development~\cite{Pe, ZN, Mu, GSS}.

As is well known, under the assumptions of spatial homogeneity and isotropy, Newtonian cosmology is  an excellent approximation to the standard FLRW cosmological models of general relativity so long as the net pressure, as a source of gravity, can be neglected in comparison with the energy density of the matter content of the universe. More generally, it turns out that after recombination, Newtonian gravitation can be applied in the study of nonrelativistic motion of matter on subhorizon scales~\cite{Pe, ZN, Mu, GSS}. 

In this paper, following the familiar approach of Newtonian cosmology, we consider the dynamics of a large gas cloud in accordance with  the Newtonian regime of nonlocal gravity. Can nonlocal gravity solve the problem of structure formation in cosmology? An adequate treatment of this subject remains a task for the future; however, we make a beginning here toward the extension of the Newtonian regime of nonlocal gravity to the cosmological domain. 

In standard Newtonian cosmology, the work of Zel'dovich on the nonlinear growth of perturbations has resulted in a useful model for the large scale structure of the universe~\cite{Pe, ZN, Mu}; for a recent review of this topic, see Ref.~\cite{GSS}.  The Zel'dovich solution is treated in Ref.~\cite{Mu} and the connection of the Zel'dovich approach to models of the cosmic web is discussed in Refs.~\cite{Pe},~\cite{Mu} and~\cite{GSS}. In this paper, we formulate the nonlocal analog of the Zel'dovich solution that involves, in effect, only one spatial dimension~\cite{Mu}. The resulting nonlocal and nonlinear integro-differential equation and its solutions  are then studied. Moreover, we formulate and implement a numerical algorithm for approximating the  solution of this system.  In conformity with the interpretation of the Zel'dovich solution~\cite{Mu, GSS}, we are interested in the growth of gravitational instability on the largest scales. Can nonlocal gravity bring about large scale structure without recourse to any dark matter? This is the basic question that is addressed in the present work. In this first attempt at a treatment of an aspect of this fundamental problem, our results indicate that nonlocal gravity has the potential to solve the problem of structure formation in cosmology.

\section{Nonlocal Kernel}

According to nonlocal gravity theory~\cite{NL1, NL2, NL3, NL4, CM, NL5}, nonlocality persists in the Newtonian regime. Indeed, Poisson's equation of Newtonian theory of gravitation is linearly modified as follows
\begin{equation}\label{I0}
\nabla^2 \phi(x, t) + \int\kappa(x-y, t)\,\nabla^2 \phi(y, t)\,d^3y =4\pi G \,\rho(x, t)\,,
\end{equation}
where $\phi$ is the gravitational potential and $\kappa$ is the \emph{convolution kernel} of nonlocal gravity in the Newtonian regime. As discussed in detail in Ref.~\cite{NL7}, under reasonable mathematical conditions, it is possible to write Eq.~\eqref{I0} as 
\begin{equation}\label{I1}
\nabla^2 \phi=4\pi G \,(\rho+ \rho_{D})\,,
\end{equation}
where  the nonlocal aspect of the gravitational interaction simulates dark matter with an effective density $ \rho_D$ given by
\begin{equation}\label{I2}
 \rho_D (x, t)=\int {\cal Q}(x-y, t)\, \rho(y, t)\,d^3y\,.
\end{equation}
That is, the density of the \emph{effective} dark matter in this theory is given by the convolution of the matter density with the \emph{reciprocal kernel} ${\cal Q}$ of nonlocal gravity. Indeed, $\kappa$ 
and $\mathcal{Q}$ are reciprocal to each other and satisfy the reciprocity relation
\begin{equation}\label{I2a}
 \kappa (x-y, t) + \mathcal{Q}(x-y, t) + \int  \kappa (x-z, t) \, \mathcal{Q}(z-y, t)\, d^3z=0\,,
\end{equation}
which is symmetric with respect to the interchange of  $\kappa$ and $\mathcal{Q}$ and follows from Eqs.~\eqref{I0}--\eqref{I2}. Furthermore, in the spatial Fourier domain, Eq.~\eqref{I2a} implies that 
\begin{equation}\label{I2b}
 \hat{\kappa} (k, t)=-\frac{\hat{\mathcal{Q}}(k, t)}{\hat{\mathcal{Q}}(k, t)+1}\,,
\end{equation}
provided  $\hat{\mathcal{Q}}(k, t)+1\ne 0$~\cite{NL7}. 

It is natural to compare Eq.~\eqref{I1} with the observational data regarding dark matter in astrophysics~\cite{RF, RW, SR}. At the present stage of the development of the theory, the actual functional form of the reciprocal kernel of nonlocal gravity ${\cal Q}$ must be determined from observational data. Once the reciprocal kernel ${\cal Q}$ is known, the kernel of nonlocal gravity $\kappa$ can be determined provided  that  ${\cal Q}$ is a function that is absolutely integrable ($L^1$) as well as square integrable ($L^2$) over all space; moreover, its spatial Fourier transform, ${\hat {\cal Q}}(k, t)$, should be such that  ${\hat {\cal Q}}(k, t) +1 \ne 0$~\cite{NL7}. We now proceed to the determination of  the reciprocal kernel ${\cal Q}$ from current astrophysical data.

\subsection{Reciprocal Kernel $q$ at the Present Epoch}

Consider the circular motion of stars and gas clouds in the disk of a spiral galaxy. According to the kinematics of uniform circular motion, the centripetal acceleration involved in such motion is $v_0^2/|r|$, where $v_0$ is the uniform circular speed and $r$ is the radius vector from the center of the galaxy to the star or gas cloud such that $|r|$ ranges from the exterior of the central massive bulge of mass $M$ to the outer reaches of the galaxy. Detailed observations of the rotation curves of spiral galaxies indicate that $v_0$ is nearly constant for the stars and gas clouds under consideration~\cite{RF, RW, SR}.  Newton's second law of motion then implies that in the region under consideration here  the force of gravity must decrease with distance as $1/|r|$. This significant deviation from the inverse square force law can be attributed to the existence of  ``dark matter" with an effective  density of  $\rho_D$.  The modified Poisson Eq.~\eqref{I1} is linear; therefore, the gravitational potential $\phi$ is the sum of contributions due to $\rho$ and $\rho_D$. Thus, assuming spherical symmetry, $\nabla \phi_D=v_0^2\, \hat{r}/|r|$, where $\phi_D$ is the gravitational potential due to $\rho_D$ and $\hat{r}$ is the radial unit vector.  It follows that $\nabla^2 \phi_D=v_0^2/|r|^2$, and hence $\rho_D=v_0^2/(4 \pi G \,|r|^2)$. Assuming that the galactic core acts on the stars and gas clouds under consideration here as though its mass were concentrated at its center, we can write $\rho(x, t) = M\, \delta(x)$, where $M$ is the mass contained in the nuclear bulge. Then the solution of Eq.~\eqref{I1} can be written as 
\begin{equation}\label{I3}
\phi_T=-\frac{GM}{|r|}
+\frac{GM}{\bar{\lambda}}\ln\left(\frac{|r|}{\bar{\lambda}}\right)\,.
\end{equation}
where $\bar{\lambda}=GM/v_0^2$ is expected to be a universal galactic length of  order 1\,kpc $\approx 3.1 \times 10^{21}$ cm. Moreover, we find from Eq.~\eqref{I2} that the reciprocal kernel in this case is given by
\begin{equation}\label{I4}
q_K=\frac{1}{4\pi\bar{\lambda}}\,\frac{1}{|x-y|^2}\,.
\end{equation}

It is noteworthy that  in this way we recover in the Newtonian regime of nonlocal gravity  the phenomenological Tohline--Kuhn scheme of modified gravity~\cite{T1,T2, K1, K2}. Indeed, Tohline suggested in 1983 that the gravitational potential of a point mass $M$ could be modified as in Eq.~\eqref{I3} in order to account for galactic dynamics without recourse to dark matter~\cite{T1, T2}. Tohline's work was later generalized by Kuhn and his collaborators by postulating Eqs.~\eqref{I1} and~\eqref{I2} with the Kuhn kernel~\eqref{I4}~\cite{K1,K2}. A lucid account of the Tohline--Kuhn approach is contained in the review paper of Bekenstein~\cite{B}. Nonlocal gravity places the Tohline--Kuhn scheme in a fully relativistic framework. 

The Kuhn kernel accounts for the observational data regarding the \emph{flat} rotation curves of spiral galaxies; therefore, its range of validity extends from the radius of the nuclear core to the outer reaches of a spiral galaxy. To come up with a reciprocal kernel $q$ that is valid over all space, we need to extend the Kuhn kernel $q_K=|r|^{-2}/(4\pi\,\bar{\lambda})$ for $|r| \to \infty$ and $|r| \to 0$; moreover, the resulting kernel should be $L^1$, $L^2$ and $\hat{q} +1 \ne 0$. 
 
The behavior of the kernel for $|r| \to \infty$ is governed by the fading of spatial gravitational memory. The simplest assumption involves an exponential decay of spatial memory, $\exp{(-\mu_0\,|r|)}$, where $\mu_0^{-1}$ is a characteristic constant length over which the memory decays by a factor of $1/e$. Let us therefore consider the function $q_K\,\exp{(-\mu_0\,|r|)}$, where for $|r| \ll \mu_0^{-1}$, we recover the Kuhn kernel modified by a term linear in $\mu_0\,|r|$. To cancel this linear term and hence obtain a better approximation to the nearly flat rotation curves of spiral galaxies, we consider instead
$q_K\,(1+\mu_0\,|r|)\,\exp{(-\mu_0\,|r|)}$. This function is $L^1$, but \emph{not} $L^2$ due to the singularity of the Kuhn kernel at $r=0$. The simplest way to moderate this singularity is to let $|r| \to a_0+ |r|$, where $a_0$ is a new constant short-range length parameter. In this way, we finally arrive at two possible $L^2$ functions~\cite{NL7} 
\begin{equation}\label{I4a}
q_1=\frac{1}{4\pi \lambda_0}~ \frac{1+\mu_0 (a_0+|r|)}{|r|\,(a_0+|r|)}~e^{-\mu_0 |r|}\,
\end{equation}
and
\begin{equation}\label{I4b}
q_2=\frac{1}{4\pi \lambda_0}~ \frac{1+\mu_0 (a_0+|r|)}{(a_0+|r|)^2}~e^{-\mu_0 |r|}\,,
\end{equation}
where we have defined 
\begin{equation}\label{I4c}
\frac{1}{\lambda_0} :=\frac{1}{\bar{\lambda}}\,e^{-\mu_0\,a_0}\,.
\end{equation}

The spatial Fourier transforms of $q_1$ and $q_2$ have been studied in detail in Ref.~\cite{NL7}. It turns out that $\hat{q}_1>0$ and the corresponding nonlocal kernel $\kappa_1$ can be obtained explicitly  from Eq.~\eqref{I2b} in this case. For $q_2$, however, it turns out that $\hat{q}_2>-a_0 / \lambda_0$, so that  ${\hat {q}}_2 +1 \ne 0$ provided that $a_0 < \lambda_0$; in this case, $\kappa_2$ can be obtained explicitly  from the corresponding Eq.~\eqref{I2b} as well. 

Starting from the Kuhn kernel~\cite{K1, B}, two possible functional forms for the reciprocal kernel have been obtained, namely, $q_1$ and $q_2$. Each kernel contains three positive length parameters  $\lambda_0$, $a_0$ and $\mu_0^{-1}$, where $\lambda_0$ is related to the main (Tohline--Kuhn) nonlocality parameter of order 1 kpc, while $a_0$ and $\mu_0$ modify the behavior of the Kuhn kernel for  $r \to 0$ and $r \to \infty$, respectively. The exponential decay of these kernels with increasing distance is a reflection of the fading of spatial memory.  If $a_0=0$ and $\mu_0=0$ in Eqs.~\eqref{I4a}--\eqref{I4c}, $q_1$ and $q_2$ both reduce to the Kuhn kernel 
$|r|^{-2}/(4\pi\,\bar{\lambda})$. It remains to compare the theory with observational data. 

For applications to the rotation curves of spiral galaxies and the internal dynamics of clusters of galaxies, we may neglect the short-range parameter $a_0$~\cite{NL6}.  With $a_0=0$, $q_1$ and $q_2$  both reduce to $q_0$,
\begin{equation}\label{I5}
 q_0=\frac{1}{4 \pi \lambda_0}\frac{(1+\mu_0 |r|)}{|r|^2}e^{-\mu_0 |r|}\,.
\end{equation}
Let us introduce the dimensionless parameter $\alpha_0$,
\begin{equation}\label{I6}
 \alpha_0 :=\int q_0\,d^3r = \frac{2}{\lambda_0 \mu_0}\,,
\end{equation}
so that instead of $\lambda_0$ and $\mu_0$, kernel $q_0$ can be characterized by the dimensionless parameter $\alpha_0$ and the memory (``Yukawa") parameter $\mu_0$. It follows from an extensive analysis of astrophysical data that~\cite{NL6} 
\begin{equation}\label{I7}
\alpha_0 = 10.94\pm2.56\,, \qquad  \mu_0 = 0.059\pm0.028~{\rm kpc^{-1}}\,.
\end{equation}
Therefore,  $\lambda_0=2/(\alpha_0\, \mu_0)$ works out to be $\lambda_0 \approx 3 \pm 2~ {\rm kpc}$. 

The determination of the short-range parameter $a_0$ has been discussed in Ref.~\cite{ChMa}. Based on the behavior of nonlocal gravity in the Solar System and the current data regarding the precession of the perihelia of the planetary orbits, it is possible to establish lower limits on $a_0$. The observational data for the perihelion precession of Saturn, for instance,  can be used to set a preliminary lower limit of  $a_0 \gtrsim 2 \times 10^{15}$  cm  for $q_1$ and   $a_0 \gtrsim 5.5 \times10^{14}$ cm for $q_2$~\cite{ChMa}. 
Eventually, it should be possible to determine $a_0$ and choose either $q_1$ and $q_2$ to represent the nonlocal kernel in the Newtonian regime. 

In nonlocal gravity, the kernel must ultimately be determined from the observational data. The simple universal three-parameter functional forms presented above are clearly not unique; in fact, more complicated expressions including other parameters cannot be excluded. In this connection, it is important to remark that an acceptable theory of gravitation must agree with Newton's theory in some limit. In the \emph{Principia}, Newton investigated the physical consequences of various functional forms for the gravitational force such as $r$ and $r^{-3}$; however, he chose $r^{-2}$, as this case uniquely  led to Kepler's empirical laws of planetary motion~\cite{Cohen}. Moreover, observational data have limited accuracy; therefore, to the inverse square force law, for instance, one can add other functional forms with coefficients so small as to be consistent with available experimental results.

\subsection{Reciprocal Kernel $\mathcal{Q}(r, t)$ in Cosmology}

Nonlocal gravity, which is the classical nonlocal extension of general relativity,  has only been studied in the linear weak-field approximation~\cite{NL5}. Nothing is known about strong-field situations, such as  black holes; in particular, no cosmological solution of nonlocal gravity is known.  On the other hand, in certain situations, Newtonian cosmology provides a good approximation to the homogeneous and isotropic FLRW models of relativistic cosmology.  Indeed, Milne and McCrea showed in 1934 that the dynamics of the universe given by Newtonian cosmology is the same as in the standard general relativistic FLRW cosmological models so long as pressure can be neglected~\cite{Mi, Mc}.

To proceed, we may tentatively assume that an extension of the Newtonian regime of nonlocal gravity to the cosmological domain could be useful. To this end, an appropriate reciprocal kernel should properly incorporate the expansion of the universe. The kernel must reflect the gravitational memory of past history of the universe,  satisfy the mathematical requirements of nonlocal gravity~\cite{NL7} and reduce to the currently accepted kernel $q$ at the present epoch $t=t_0$. Memory dies out in time and space.  The fading of memory in time implies that in nonlocal gravity  the strength of the gravitational interaction must  decrease with cosmic time. Therefore, we will assume in this paper that the kernel of \emph{nonlocal Newtonian cosmology} is given by
\begin{equation}\label{I8}
\mathcal{Q}(r, t)= \frac{q}{B(t)}\,,
\end{equation}
where $q$ is the kernel of nonlocal gravity at the present epoch $t=t_0$, and $B(t)$ is a monotonically increasing function of cosmic time and,  at the present epoch,  $B(t_0)=1$. 

Cosmology deals with the large scale structure of the universe; therefore, the short-range parameter $a_0$ may be deemed irrelevant in the cosmological context.  With $a_0=0$, we assume that the kernel of nonlocal Newtonian cosmology has the same spatial form as $q_0$, namely,
\begin{equation}\label{I9}
q= \frac{1}{4\pi \lambda_0}\, \frac{1+\mu_0\,|r|}{|r|^2}\,e^{-\mu_0\,|r|}\,.
\end{equation}
It is interesting to consider the density of effective dark matter to baryonic matter, $\rho_D/\rho$, for a spatially homogeneous model of matter density $\rho(t)$, which is naturally assumed  throughout  to be largely baryonic.  We find from Eq.~\eqref{I2}  that 
\begin{equation}\label{I10}
 \frac{\rho_D}{\rho}=  \int_{\mathbb{R}^3} \mathcal{Q}(r, t)\,d^3r := \alpha (t)\,,
\end{equation}
where Eqs.~\eqref{I6},~\eqref{I8} and~\eqref{I9} imply
\begin{equation}\label{I11}
\alpha (t)=\frac{\alpha_0}{B(t)}\,
\end{equation}
and $ \alpha_0=2/(\lambda_0\, \mu_0)$. It remains to specify the function $B(t)$.

In an astrophysical system, let the dark matter fraction $f_{DM}$ denote the ratio of the total mass of dark matter to the total mass of baryonic  matter in the system. Combining Eqs.~\eqref{I10} and~\eqref{I11}, we see that in a uniform density cosmological model the effective dark matter fraction $f_{DM}=\alpha(t)$ \emph{decreases} with cosmic time. Indeed, it is natural to assume that relative to baryonic matter,  the effective dark matter  was more plentiful in the past, since nonlocality is connected with the memory of the past state of the gravitational field and memory fades with time. 
It is important to compare this aspect of our toy model with the currently accepted model of standard cosmology, where $f_{DM}$ for the universe is about 5 and independent of cosmic time. Does $f_{DM}$ actually evolve with cosmic time, or equivalently, with the cosmological redshift $z$? It appears that this empirical question has not yet been tackled by observational cosmologists, as such an issue does not even arise in the current paradigm of dark matter. 

Henceforward, we will assume for the sake of definiteness  that 
\begin{equation}\label{I12}
 \lambda_0 \approx 3\, {\rm kpc}\,, \qquad \frac{1}{\mu_0} \approx 17\, {\rm kpc}\,, \qquad  \alpha_0 \approx 11\,.
\end{equation}
A toy model of nonlocal Newtonian cosmology, which will turn out to be rather  similar to that of the standard Milne--McCrea Newtonian cosmology, is introduced in the next section.

\section{Nonlocal Newtonian Cosmological Model}

The cosmological Euler--Poisson model combines conservation of mass and momentum with the nonlocally modified Newtonian gravitational potential:
\begin{align}
\rho_t+\nabla\cdot (\rho v)&=0,\label{cm:cma}\\
v_t+(v\cdot \nabla) v&=-\nabla\phi-\frac{1}{\rho}\nabla P,\label{cm:cmo}\\
\nabla^2 \phi&=4\pi G(\rho+\rho_D),\label{cm:pe}
\end{align}
where $\rho$ is the density of baryonic matter, $v$ is its velocity field, $P$ is the pressure and $\phi$ is the gravitational potential. Moreover,  $\rho_D$ is  the density of effective dark matter given by 
\begin{equation}\label{dark}
 \rho_D (x, t)=\int \mathcal{Q}(x-y, t)\, \rho(y, t)\,d^3y\,,
\end{equation}
where 
\begin{equation}\label{kernel}
\mathcal{Q}(r, t)= \frac{\alpha_0\,\mu_0}{8\pi\,B(t)}\, \frac{1+\mu_0\,|r|}{|r|^2}\,e^{-\mu_0\,|r|}\,.
\end{equation}
We note that subscript $t$ denotes partial derivative with respect to cosmic time; e.g.,  $\rho_t=\partial \rho/\partial t$, etc. 

To obtain the nonlocal analog of the standard models of Newtonian cosmology, we assume an infinite  spatially homogeneous and isotropic perfect fluid medium with $\rho=\bar{\rho}(t)$ and $P=P(\bar{\rho})$ that is expanding uniformly. This expansion of the universe can be expressed via $r=A(t)\, \xi$, where $A(t)$ is the scale factor, $r$ denotes the spatial position of the fluid particle at time $t$ and $\xi$ denotes the spatial position of the particle at some fiducial time $t_0$ such that $A(t_0)=1$. We take $t_0$ to be the present epoch. Thus, $\bar{v}=dr/dt = H(t)\, r$, where $H(t)=\dot A /A$ is the Hubble parameter and an overdot denotes differentiation with respect to time. Moreover, it follows from Eqs.~\eqref{I10} and~\eqref{I11} that for an infinite homogeneous matter distribution $\rho_D=\alpha(t) \rho$, where $\alpha(t)=\alpha_0/B(t)$. 
The explicit solution of the  system~\eqref{cm:cma}--\eqref{cm:pe} is thus given by
\begin{align}
\label{cm:bexsol}
\nonumber\bar{\rho}&= A^{-3} \rho_0, \qquad P=P(\bar{\rho})\,, \\
\nonumber \bar{v}&=\dot A A^{-1} r,\\
\bar{\phi}&=-\frac{1}{2} \ddot A A^{-1} |r|^2,
\end{align}
where $\bar{\phi}$ is determined up to an integration constant and the scale factor $A$ is a solution of the differential equation
 \begin{equation}\label{cm:ode2}
A^{-1}\ddot A =-\frac{4\pi G}{3} A^{-3}(1+\alpha)\rho_0\,.
\end{equation}
For the initial data 
\begin{equation}
\label{cm:ode1IC}
A(t_0)=1, \qquad \dot A(t_0)=H_0
\end{equation}
prescribed at the present epoch $t_0$ using the Hubble constant $H_0$,  we obtain the Hubble flow for this model provided $B(t)$ is known.  A remark is in order here regarding the necessity of a spatially infinite cosmological solution:  For a spherically symmetric matter distribution of finite radius, a uniformly expanding homogeneous distribution is possible only if the standard Poisson equation of Newtonian gravity is valid, i.e., Newton's inverse square law of gravitation is maintained. 

To go forward, we must assume a functional form for $B(t)$; for example, $B(t)$ could be  $A^{\varpi}$, where $\varpi>0$ is a constant.  For an expanding universe where the effective dark matter fraction decreases as the universe expands, perhaps the  simplest model---the one that we will discuss---is derived from the assumption
\begin{equation}
\label{cm:newa}
B(t)=A(t)\,, \qquad  \alpha=\frac{\alpha_0}{A}\,,
\end{equation}
where  $\alpha_0\approx 11$ is the proportionality constant at the present age ($t_0$) of the universe.  The differential equation~\eqref{cm:ode2} then takes the form
 \begin{equation}\label{cm:ode3}
A^{-1}\ddot A =-\frac{4\pi G\rho_0}{3} A^{-3} \left(1+\frac{\alpha_0}{A}\right)\,.
\end{equation}

Equation~\eqref{cm:ode3} can be integrated once to obtain
\begin{equation}\label{cm:ode4}
\frac{1}{2} \dot A^2= \frac{4\pi G\rho_0}{3} \left(\frac{1}{A}+\frac{\alpha_0}{2 A^{2} }\right)+E,
\end{equation}
where the total energy parameter $E$ is the constant of integration. We set $E = 0$, just as in  the critical case in standard Newtonian cosmology, which corresponds to the spatially flat FLRW universe.  

For this analog of the flat FLRW universe, a second integration yields the equation
\begin{equation}\label{cm:ode4sol}
\frac{1}{3} (2 A+\alpha_0)^{3/2}-\alpha_0 (2 A+\alpha_0)^{1/2} = 2\, \left(\frac{4 \pi G\rho_0}{3} \right)^{1/2}\,t+C
\end{equation}
for a new integration constant $C$. Only positive square roots are considered throughout. 
Assuming that the scale factor vanishes at the Big Bang, $t=0$, $C$ can be evaluated from $A(0)=0$ and the formula for $A$ thus reduces to 
\begin{equation}\label{cm:ode5}
\frac{1}{3} (2 A+\alpha_0)^{3/2}-\alpha_0 \,(2 A+\alpha_0)^{1/2} = 2 \left(\frac{4 \pi G\rho_0}{3} \right)^{1/2}\, t-\frac{2}{3}\,\alpha_0\sqrt{\alpha_0}\,.
\end{equation}
The unknown $A(t)$ can be obtained by solving for the positive real root of the corresponding cubic equation for $\sqrt{2 A+\alpha_0}$.

The age of the universe $t_0$, determined by $A(t_0)=1$ in our model, is 
\begin{equation}\label{cm:aou}
t_0= \frac{ \alpha_0\sqrt{\alpha_0}-(\alpha_0-1) \sqrt{\alpha_0+2} } {\sqrt{12\pi G\rho_0}}\,.
\end{equation}
It is interesting to note that the numerator of this expression for $t_0$ is a positive and monotonically decreasing function of $\alpha_0$; in fact, it starts from $\sqrt{2}$ at $\alpha_0=0$ and vanishes asymptotically as $\alpha_0 \to \infty$. Moreover, the present value of the Hubble parameter is given by 
Eq.~\eqref{cm:ode4} in this model with $E = 0$, namely, 
\begin{equation}\label{II1}
H_0^2=\frac{4 \pi G \rho_0}{3}\, (\alpha_0 +2)\,,
\end{equation}
so that 
\begin{equation}\label{II2}
3\,H_0\, t_0=\alpha_0\,\sqrt{\alpha_0\,(\alpha_0 +2)}-(\alpha_0-1)\, (\alpha_0 +2)\,.
\end{equation}
For $t/t_0 \to \infty$, it follows from Eq.~\eqref{cm:ode5} that asymptotically $A^3 \sim 6 \pi G \rho_0\, t^2$, so that this model approaches the Newtonian analog of the Einstein--de Sitter model as $t/t_0 \to \infty$.  

The uniformly expanding model of nonlocal Newtonian cosmology under consideration here reduces to the Newtonian version of the spatially flat Einstein--de Sitter model if we formally let $\alpha_0 \to 0$. Indeed, with $\alpha_0 = 0$, the work throughout this  paper reduces to the standard \emph{local} treatment of gravitational instability in Newtonian cosmology~\cite{Mu}. 

With a Hubble constant of $H_0 \approx 70$ km s$^{-1}$ Mpc$^{-1}$, we find that,  according to the nonlocal  model with $\alpha_0\approx 11$, the present density of baryons in the universe is $\rho_0 \approx 1.4\times  10^{-30}$ g \,cm$^{-3}$ and the age of the universe is $t_0 \approx 7.21 \times 10^9$ yr.  Thus our  model gives a value  for the age of the universe that is about half of the age of the currently accepted model; moreover, the currently accepted baryon density is about $0.3\,\rho_0$~\cite{GSS}. These cosmological parameters are somewhat reminiscent of the corresponding parameters for the spatially flat Einstein--de Sitter model of standard cosmology. 

In the present paper, we employ this \emph{toy} model of nonlocal Newtonian cosmology  in order to explore the possibility that large scale structure formation in cosmology may be possible without invoking dark matter. The next section contains an analysis of the \emph{linear} stability of our model. We then discuss the \emph{nonlinear} instability of the model using the Zel'dovich approach.

\section{Jeans Instability}

The instability of self-gravitating static configurations was first studied by Jeans within the framework of Newtonian  theory of gravitation~\cite{Je, Jea}. Bonnor extended the study of Newtonian gravitational instability to expanding homogeneous media~\cite{Bo}. In the context of modern cosmology, gravitational instability has been treated, for instance, in Refs.~\cite{Pe, ZN, Mu}.  To investigate Jeans instability in our nonlocal model,  we devote this section to a \emph{linear} perturbation analysis of the model away from the  homogeneous and isotropic  solution given in Eq.~\eqref{cm:bexsol}. The perturbed solution has a baryonic density given by $\bar \rho + \delta \rho$, where $ \bar \rho = \rho_0/A^3$ is the background baryonic density and $\delta{\rho}/\bar{\rho}$ can be expressed as a sum of Fourier modes each with wave vector $k$, 
\begin{equation}\label{3.1}
\frac{\delta \rho}{\bar \rho}=\varepsilon_0 \sum_k D_k(t) e^{ik \cdot r/A(t)}\,,
\end{equation}
where $\varepsilon_0$, $0<\varepsilon_0 \ll 1$, is a constant perturbation parameter and we have used the fact that the spatial scale of the perturbation expands with the universe. The perturbed flow velocity is given by $\bar v+\delta v$, where $ \bar v=(\dot A/A)\,r$ and 
\begin{equation}\label{3.2}
\delta v=\varepsilon_0 \sum_k V_k(t) e^{ik \cdot r/A(t)}\,.
\end{equation}
For the fluid pressure, $P=P(\rho)$, we find that the net perturbed pressure is
\begin{equation}\label{3.3}
P(\bar{\rho} + \delta \rho)=P(\bar{\rho})+ \left . \frac{dP}{d\rho}\right |_{\rho\,=\,\bar{\rho}}\, \delta \rho
\end{equation}
by Taylor expansion.  We recall that the speed of sound in the medium is given by 
\begin{equation}\label{3.4}
c_s (\rho)=\sqrt{\frac{dP}{d\rho}}\,.
\end{equation}

We are interested here in the linear \emph{adiabatic} perturbation of our perfect fluid model; therefore, in the substitution of perturbed values of the fluid parameters in the model 
Eqs.~\eqref{cm:cma}--\eqref{cm:pe}, we keep terms only up to first order in $\varepsilon_0$. As a direct result of this linearity, it is possible to use complex perturbation amplitudes for the sake of simplicity, with the proviso that only the real parts of the perturbed equations have physical significance.  
\begin{figure}
\centerline{\includegraphics[width=4in]{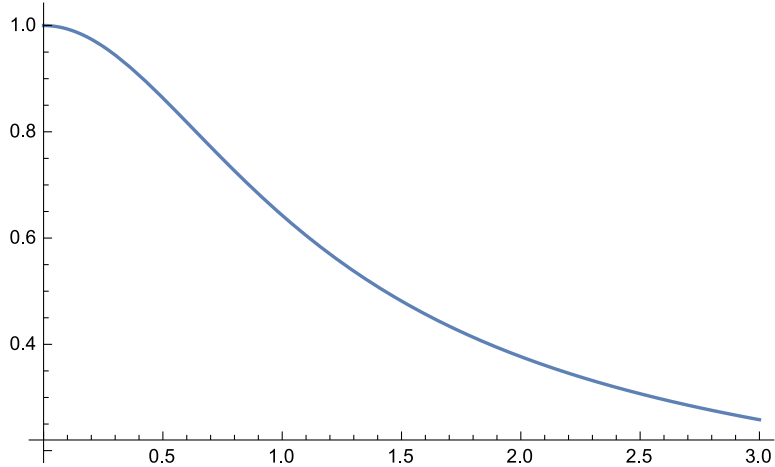}}
\caption{Graph of $\frac{1}{2} [ \Theta^{-1} \arctan\Theta+(1+\Theta^2)^{-1}]$ versus $\Theta$.   \label{fig: fig1}}
\end{figure} 

It follows from the perturbed continuity equation that 
\begin{equation}\label{3.5}
A \dot D_k + i \, k \cdot V_k=0\,.
\end{equation}
Moreover, the perturbed Euler equation can be written as
\begin{equation}\label{3.6}
-\nabla \phi= \frac{\ddot A}{A}\,r + \varepsilon_0 \sum_k \big(\dot V_k + \frac{\dot A}{A}\, V_k +i\, \frac{\bar{c}_s^{\,2}}{A}\, k \, D_k\big)\,e^{ik \cdot r/A(t)}\,,
\end{equation}
where $\bar{c}_s := c_s (\bar{\rho})$. The divergence of Eq.~\eqref{3.6} is given by
\begin{equation}\label{3.7}
-\nabla^2 \phi=3\, \frac{\ddot A}{A}\, + \varepsilon_0 \sum_k \big[\frac{1}{A}\,(i\, k\cdot \dot V_k)+\frac{\dot A}{A^2} \,(i\, k\cdot  V_k) - \frac{\bar{c}_s^{\,2}}{A^2}\, k^2 \, D_k\big]\,e^{ik \cdot r/A(t)}\,.
\end{equation}
To simplify matters, let us note that Eq.~\eqref{3.5} implies, upon differentiation with respect to time, that
\begin{equation}\label{3.8}
\dot A \dot D_k +A \ddot D_k + i \, k \cdot \dot V_k=0\,.
\end{equation}
Thus substituting for $i \, k \cdot V_k$ and $i \, k \cdot \dot V_k$ from Eqs.~\eqref{3.5} and~\eqref{3.8}, respectively, in Eq.~\eqref{3.7}, we find
\begin{equation}\label{3.9}
\nabla^2 \phi=-3\, \frac{\ddot A}{A}\, + \varepsilon_0 \sum_k \big(\ddot D_k +2\,\frac{\dot A}{A} \,\dot D_k  + \frac{\bar{c}_s^{\,2}}{A^2}\, k^2 \, D_k\big)\,e^{ik \cdot r/A(t)}\,.
\end{equation}
Finally, using this expression in the modified Poisson Eq.~\eqref{cm:pe} and utilizing 
Eq.~\eqref{cm:ode2}, we obtain
\begin{equation}\label{3.10}
\frac{d^2D_k}{dt^2}+ 2\, \frac{\dot A}{A}\,\frac{dD_k}{dt} + {\cal J}\, D_k =0\,,
\end{equation}
where ${\cal J}$ is given by
\begin{equation}\label{3.11}
{\cal J}=\frac{k^2\,\bar{c}_s^{\,2}}{A^2}- 4 \pi G \bar \rho \, [1+Q_k(t)]\,.
\end{equation}
Here $Q_k$ is the spatial Fourier integral transform of $\mathcal{Q}$, namely, 
\begin{equation}\label{3.12}
Q_k(t) := \int \mathcal{Q}(r, t)\, e^{-ik \cdot r/A(t)}\, d^3r\,,
\end{equation}
where the reciprocal kernel is
\begin{equation}\label{3.12a}
\mathcal{Q}(r, t)= \frac{\alpha_0\, \mu_0}{8\pi\,A(t)}\, \frac{1+\mu_0\,|r|}{|r|^2}\,e^{-\mu_0\,|r|}\,.
\end{equation}
Indeed, it is possible to show that 
\begin{equation}\label{3.13}
Q_k(t) :=\frac{\alpha_0}{2\, A(t)}\left[\frac{1}{\Theta}\, \arctan{\Theta} +\frac{1}{1+\Theta^2}\right]\,,\end{equation}
where
\begin{equation}\label{3.14}
\Theta :=\frac{|k|}{\mu_0\,A(t)}\,.
\end{equation}
Thus $Q_k(t)$ depends only on the wave number $|k|$ and cosmic time $t$. As $|k|/\mu_0 \to 0$, we find $Q_0(t)=\alpha_0/A(t) = \alpha(t)$, while for $|k|/\mu_0 \to \infty$, we have $Q_{\infty} =0$, see Figure 1. 

Our linear perturbation analysis makes it possible to consider $\delta \rho$ and $\delta v$ as superpositions of different Fourier modes characterized by the wave vector $k$. Suppose, for instance, that for a mode $k$, ${\cal J}$ is nonzero and the corresponding peculiar velocity is orthogonal to $k$; that is, $k\cdot V_k=0$. Then, it follows from the perturbation equations that there is no change in density for this \emph{vector} mode, since $D_k=0$. On the other hand, suppose that $V_k$ is parallel to $k$, then $D_k$ is in general nonzero and is given by Eq.~\eqref{3.10}.

\begin{figure}
\centerline{\includegraphics[width=4in]{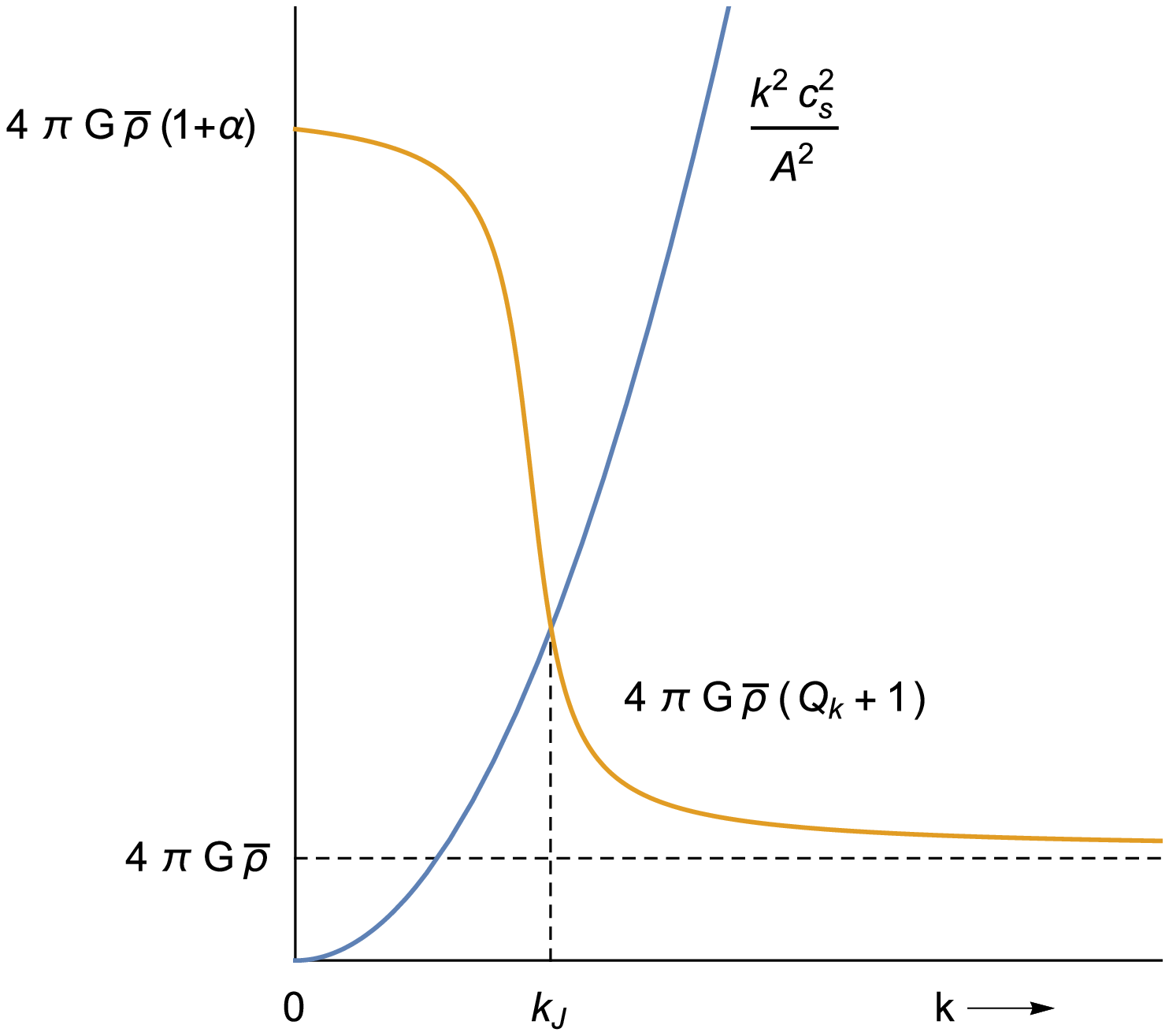}}
\caption{Schematic plot that illustrates the determination of Jeans length, $\lambda_J=2\,\pi\, A(t)/k_J$,  in our model. Here $k$,  $k :0 \to \infty$,  is the wave number.\label{fig: fig2}}
\end{figure}

The solutions of Eq.~\eqref{3.10} are oscillatory sound waves for ${\cal J}>0$, while for ${\cal J} <0$,  we have a linear combination of growing and decaying modes. The transition between the two regimes is characterized by ${\cal J}=0$, which can be solved for wave number $|k|$. The result, $k_J$, can be written as
\begin{equation}\label{3.15}
k_J =\frac{2\,\pi\,A(t)}{\lambda_J }\,,
\end{equation}
which defines the \emph{Jeans length} $\lambda_J$ given by 
\begin{equation}\label{3.16}
\frac{\lambda_J}{\bar{c}_s}=\Big[\frac{\pi}{G\bar \rho (1+Q_k)}\Big]^{1/2}\,.
\end{equation}
The determination of $k_J$ via $\mathcal{J}=0$ is illustrated in Figure 2. 

The attraction of gravity tends to produce clumps; however, pressure forces work against this tendency. In Eq.~\eqref{3.16}, which expresses the Jeans criterion, these forces balance each other such that  $\lambda_J/\bar{c}_s$, which is the time that it would take for a pressure wave to move across a Jeans length, is comparable to the gravitational response time of a self-gravitating fluid of density $\bar \rho$. For $\lambda \ll \lambda_J$, or $|k| \gg k_J$, the density contrast oscillates as a sound wave, while for 
$\lambda \gg \lambda_J$, or $|k| \ll k_J$, the pressure term can be neglected and gravitational instability takes over. This is due to the fact that the gravitational response time is very short in comparison with the period of the pressure wave.

We are interested in the study of \emph{nonlinear} gravitational instability in our model. Henceforth, we neglect the pressure term in the Euler Eq.~\eqref{cm:cmo}. Thus the structures of interest in the post-recombination era have extensions  that  are much larger than the corresponding Jeans length. At the recombination era, corresponding to a cosmological redshift of $z \approx 10^3$, when structures are expected to start forming, it is possible to estimate $\lambda_J $ and the corresponding Jeans mass 
$M_J:=\bar \rho \,\lambda_J ^3$. To get a rough idea of the orders of magnitude involved here, we recall that if we formally set $\alpha_0 =0$ in our nonlocal scheme, we recover the Newtonian analog of the Einstein--de Sitter model. For the standard Einstein--de Sitter model, the result is $\lambda_J \sim 20$\,pc and $M_J \sim 10^5 M_\odot$, comparable to the parameters of a globular star cluster~\cite{ZN, Pe}.

\section{Nonlocal Analog of the Zel'dovich Solution}

To study  nonlinear gravitational  instability in the context of Newtonian cosmology, Zel'dovich introduced a method that is based on Lagrangian coordinates. To appreciate his novel idea thus requires treating the nonlocal fluid model~\eqref{cm:cma}--\eqref{cm:pe} in Lagrangian coordinates~\cite{Mu}. 

When pressure is neglected as discussed in the previous section,  a solution of the fluid model must satisfy the system of equations
\begin{align}
\label{cm:prepeqsa}
\rho_t+\nabla\cdot (\rho v)&=0\,,\\
\label{cm:prepeqsb}
 \nabla\cdot [v_t+(v\cdot \nabla) v]&=-4\pi G(\rho+\rho_D)\,,
\end{align}
obtained by taking the divergence of both sides of the conservation of momentum equation  and substituting for  the Laplacian of the ``Newtonian" gravitational potential using the Poisson equation. 

This system is converted to Lagrangian coordinates $\xi$ through  the Lagrangian flow map $X$, which is defined via the flow of fluid particles in the velocity field $v$; for instance, $X(\xi,t)$ can be the solution of the initial value problem 
\begin{equation}\label{cm:velode}
\frac{dr}{dt}=v(r,t)\,, \qquad r(t_{in})=\xi\,,
\end{equation} 
where $t_{in} > 0$ is a convenient starting value, which can be some unspecified initial epoch after the Big Bang. We assume that cosmic time $t: 0 \to \infty$, where $t=0$ indicates the Big Bang. More generally, as in this paper, $\xi$ is a time independent vector that uniquely identifies a fluid particle and is constant along its path. 

A scalar function $w$ on spacetime with coordinates $(r,t)$, is transformed to  $W$ with Lagrangian variables $(\xi,t)$ via the rule
\begin{equation} \label{cm:slct}
W(\xi,t)=w(X(\xi,t),t)\,.
\end{equation}
Using the same notation, a vector function $u$ on spacetime is transformed to $U$ in Lagrangian coordinates  by
\[
DX(\xi,t)U(\xi,t)=u(X(\xi,t),t)\,,
\]
where $D$ denotes the derivative with respect to the space variables; more precisely, $D X(\xi,t)$ is the derivative of the function 
$\xi\mapsto X(\xi,t)$ evaluated at $(\xi,t)$.  The Lagrangian coordinates of a fluid particle in spacetime are constants along its path and uniquely identify the particle.  For time dependent velocity fields,  these coordinates are defined only up to the first time when two trajectories with different Lagrangian markers meet in space.  At such a  time,  the Lagrangian coordinate system breaks down and singularities called caustics are formed.  

To transform the continuity equation to Lagrangian coordinates, define 
\[w_c(r,t)=\rho_t(r,t)+\nabla \rho(r,t)\cdot v(r,t)+\rho(r,t) \nabla\cdot v(r,t)\]
 and recall Liouville's theorem 
\[
\frac{\partial}{\partial t} \det DX(\xi,t)={\rm tr}\, (Dv(X(\xi,t),t))\det DX(\xi,t)
\]
from linear systems theory (see, for example, Ref.~\cite{cccode}). Moreover,  we let $R(\xi,t):=\rho(X(\xi,t),t)$  denote the baryonic density in Lagrangian coordinates and we use formula~\eqref{cm:slct} to compute
\begin{align*}
W_c(\xi,t)&=\rho_t(X(\xi,t),t)+\nabla \rho(X(\xi,t),t)\cdot v(X(\xi,t),t)+\rho(X(\xi,t),t) \nabla\cdot v(X(\xi,t),t)\\
&=\frac{\partial}{\partial t}\rho(X(\xi,t),t)+\rho(X(\xi,t),t) \nabla \cdot v(X(\xi,t),t)\\
&=\frac{\partial}{\partial t} R(\xi,t)+R(\xi,t)\, {\rm tr}\, (Dv(X(\xi,t),t))\\
&=\frac{\partial}{\partial t}R(\xi,t)+R(\xi,t)\,[\det DX(\xi,t)]^{-1} \frac{\partial}{\partial t} \det DX(\xi,t)\,.
\end{align*}
The continuity equation states that $W_c=0$. Multiplication by  $\det DX(\xi,t)$  yields the equation 
\[0=\det DX(\xi,t) \frac{\partial}{\partial t} R(\xi,t)+R(\xi,t)  \frac{\partial}{\partial t} \det DX(\xi,t)\,.\]
With $j(\xi,t):= \det DX(\xi,t)$ and $R(\xi,t)=\rho(X(\xi,t),t)$, the continuity equation in Lagrangian coordinates in compact form is
\begin{equation}\label{celc}
\frac{\partial}{\partial t} (j R)=0\,.
\end{equation}

Using similar computations, the divergence of the momentum equation is transformed to Lagrangian coordinates.  For the main step, let 
\[ w_m(x,t):=\nabla\cdot [v_t(x,t)+( (v\cdot \nabla) v)(x,t)]\]
and compute
\begin{align*}
w_m&=(\nabla\cdot v)_t+\nabla\cdot ( v\cdot \nabla) v\\
&=\frac{\partial}{\partial t} ({\rm tr}\,(D v))+{\rm tr}\,(D(Dv\cdot v))\\
&={\rm tr}\, \frac{\partial}{\partial t} (D v)+{\rm tr}\,[D^2v(v,\cdot)+Dv^2]\\
&={\rm tr}\, [(D v)_t+D^2v(v,\cdot)]+{\rm tr}\,(Dv^2)\,.
\end{align*}
By the chain rule and the definition of $X$, 
\[
\frac{\partial}{\partial t} Dv(X(\xi,t),t)=D^2v(X(\xi,t),t)(v(X(\xi,t),t),\cdot)+Dv_t(X(\xi,t),t)\,.
\]
Using this fact, we find
\begin{align*}
W_m(\xi,t)=\frac{\partial}{\partial t} {\rm tr}\, Dv(X(\xi,t),t)+{\rm tr} Dv^2(X(\xi,t),t)\,.
\end{align*}
By Liouville's theorem 
\begin{equation}\label{pmelc}
W_m(\xi,t)=\frac{\partial}{\partial t} (\frac{\partial}{\partial t}\ln \det DX(\xi,t))+{\rm tr}\, Dv^2(X(\xi,t),t)\,,
\end{equation}
and in view of the first variational equation
\[
\frac{\partial}{\partial t} DX(\xi,t)=Dv(X(\xi,t),t)DX(\xi,t)\,,
\]
we have that
\begin{equation}\label{pmelc2}
W_m(\xi,t)=\frac{\partial}{\partial t} (\frac{\partial}{\partial t}\ln \det DX(\xi,t))+{\rm tr}\, (\frac{\partial}{\partial t}DX(\xi,t) DX(\xi,t)^{-1})^2\,.
\end{equation}
With $J(\xi,t):=DX(\xi,t)$ and the corresponding Jacobian $j(\xi,t):=\det DX(\xi,t)$, we obtain
 Eq.~\eqref{cm:prepeqsb} in Lagrangian coordinates
\begin{equation}\label{dcmlc}
\frac{\partial^2}{\partial t^2} \ln j+{\rm tr}\,\left[\left(\frac{\partial J}{\partial t} J^{-1}\right)^2\right]=-4\pi G(R+R_D)\,.
\end{equation}
Henceforth, we consider the problem of solving the Lagrangian system of Eqs.~\eqref{celc} and~\eqref{dcmlc} instead of the Eulerian system of  Eqs.~\eqref{cm:prepeqsa} and~\eqref{cm:prepeqsb}. 

To illustrate this approach, let us consider the solution of the spatially homogeneous and isotropic case~\eqref{cm:bexsol} in terms of Lagrangian coordinates. The Hubble flow in this case is in effect a simple scaling with a time dependent scale factor; therefore, we seek a Lagrangian flow map of the form 
\begin{equation}\label{blafmap}
X(\xi,t)=a(t)\, \xi\,,
\end{equation}  
where $a(t_{in})=1$. In this case, $J=a(t)$\,diag$(1,1,1)$ and $j=a^3(t)$. It follows from Eq.~\eqref{celc} that $jR=R(\xi, t_{in})$. Moreover, Eq.~\eqref{dcmlc} implies that $3\,\ddot{a}/a = -4 \pi G (1+R_D/R)\,R$. In the homogeneous case, $R$ is independent of $\xi$, $R_D/R=\alpha_0/A(t)$ and $R(t)=R(t_{in})/a^3(t)$.  We thus recover Eqs.~\eqref{cm:bexsol} and~\eqref{cm:ode3} with $a(t)=A(t)/A(t_{in})$, $R=\bar{\rho}$ and $\rho_0=R(t_{in})\,A^3(t_{in})$.

\subsection{Zel'dovich Ansatz}

To find a spatially inhomogeneous solution of Eqs.~\eqref{celc} and~\eqref{dcmlc}, we follow Zel'dovich and assume a Lagrangian flow map of the form
 \begin{equation}\label{Z1}
 X(\xi,t):=\mathbb{A}(t)\,(\xi-\mathbb{F}(\xi,t))\,,
 \end{equation}
where the general Lagrangian position vector $\xi=(\xi_1, \xi_2, \xi_3)$ is  constant along the path of a fluid particle and 
 \begin{equation}\label{Z2}
 \mathbb{F}(\xi,t):= (f(\xi_1, t), 0, 0)\,.
 \end{equation}
It follows from Eqs.~\eqref{Z1} and~\eqref{Z2} that 
\begin{equation}\label{Z3}
J=\mathbb{A}(t)\, {\rm diag}(1-\Psi, 1, 1)\,, \qquad j=\mathbb{A}^3\,(1-\Psi)\,, \qquad \Psi := \frac{\partial f}{\partial \xi_1}\,.
\end{equation}

The Lagrangian conservation of mass Eq.~\eqref{celc} implies that $jR$ is constant in time; therefore,
\begin{equation}\label{Z4}
R=\frac{R_0(\xi)}{j},
\end{equation}
where $R_0(\xi)>0$ is simply a function of the Lagrangian coordinates. Moreover, the substitution of the Zel'dovich ansatz into the Lagrangian form of the divergence of the conservation of momentum Eq.~\eqref{dcmlc}  yields, after some algebra, 
 \begin{equation}\label{Z5}
- \,3\,\frac{\ddot{\mathbb{A}}}{\mathbb{A}} 
 +2\, \frac{\dot{\mathbb{A}}}{\mathbb{A}}\, \frac{\Psi_{t}} {1-\Psi}+ \frac{\Psi_{tt}}{1-\Psi}= 4\pi G\,\Big(\frac{R_0(\xi)}{\mathbb{A}^3(1-\Psi)} +R_D\Big)\,.
 \end{equation}
We now assume, in conformity with the original Zel'dovich solution~\cite{Mu},  that
\begin{equation}\label{Z6}
\mathbb{A}=A\,, \qquad R_0(\xi) = \rho_0\,,
\end{equation}
where $A(t)$,  given by Eq.~\eqref{cm:ode3}, is the scale factor of the spatially homogeneous and isotropic background and $\rho_0$ is the uniform background baryonic density at the present epoch.  Thus, using Eqs.~\eqref{cm:ode3} and~\eqref{Z6},  Eq.~\eqref{Z5} can be written as 
 \begin{align}\label{Z7}
 \frac{\Psi_{tt}}{1-\Psi}
 +2\, \frac{\dot{A}}{A}\,\frac{\Psi_{t}} {1-\Psi}+\frac{4\pi G\rho_0 }{A^3}(1+\frac{\alpha_0}{A})=4\pi G\, \Big(\frac{\rho_0}{A^3(1-\Psi)} +R_D\Big)\,.
 \end{align}

The quantity $\Psi/(1-\Psi)$ in this context is related to the \emph{density contrast},
\begin{equation}\label{Z8}
\frac{\Psi}{1-\Psi}=\frac{R-\bar{\rho}(t)}{\bar{\rho}(t)}\,,
\end{equation} 
of baryonic matter given as a solution in the Lagrangian formulation of the model~\eqref{cm:prepeqsa}--\eqref{cm:prepeqsb} relative to the background density $\bar{\rho}(t)=\rho_0/A^3$ of the exact homogeneous solution~\eqref{cm:bexsol}. In our model, $R$ is the density of baryonic matter, $R=\rho_0/[A^3\,(1-\Psi)]$, where we assume that  $\Psi$ is a positive function that is less than unity and  the background density is bounded. The matter density in Lagrangian coordinates approaches  infinity  as  $\Psi$ approaches unity. This would indicate the breakdown of the  coordinate system. On the physical side, this circumstance can be  interpreted to mean that the presence of \emph{effective dark matter} $R_D$ in the model produces cosmic structure as time approaches this epoch.

Let us now write Eq.~\eqref{Z7} in the form 
 \begin{equation}\label{Z9}
 \Psi_{tt}+2\,\frac{\dot{A}}{A}\, \Psi_{t}-\frac{4\pi G\rho_0}{A^3}\,(1+\frac{\alpha_0}{A})\, \Psi ={\cal N}\,,
 \end{equation}
where   ${\cal N}$  is defined by
 \begin{equation}\label{Z10}
{\cal N} := \frac{4\pi G\rho_0\,\alpha_0}{A^4}\Big[\frac{A^4}{\rho_0\,\alpha_0}\,(1-\Psi)\,R_D-1\Big]\,.
 \end{equation} 
The \emph{nonlinear} part of this second-order equation for $\Psi$ is contained in $\mathcal{N}$.  It is interesting to note that this nonlinear term vanishes for $R_D=(\alpha_0/A)\, R$. In this case, $\Psi$ can be determined from the linear and homogeneous form of Eq.~\eqref{Z9} with $\mathcal{N}=0$, which coincides with the $k\to 0$ limit of the linear perturbation Eq.~\eqref{3.10} for the density contrast $D_k(t)$. This limiting case is discussed in detail in the last part of this section. 

The nonlocal Zel'dovich model under consideration here  deals with the \emph{nonlinear} perturbation $\Psi$ in the baryonic density of the spatially homogeneous and isotropic background model of nonlocal Newtonian cosmology. The background expands with scale factor $A(t)$ and $\Psi$ satisfies Eq.~\eqref{Z9}, where $\mathcal{N}$, the nonlinear part of Eq.~\eqref{Z9}, is given by Eq.~\eqref{Z10}. 

The density of the effective dark matter in the Lagrangian formulation is
\begin{equation}\label{Z11}
R_D(\xi,t)=\rho_D(X(\xi,t),t)=\rho_0 \int_{\mathbb{R}^3} \frac{\mathcal{Q}(X(\xi,t)-y,t)}{j(X^{-1}(y,t),t)}\,d^3y\,,
\end{equation}  
where the reciprocal kernel $\mathcal{Q}$ is given by Eq.~\eqref{3.12a}. The Jacobian of the transformation from the Eulerian to Lagrangian coordinates is $j$; therefore, the change of variables $y=X(\zeta,t)$ yields the more useful form of $R_D$, 
\begin{equation}\label{Z12}
R_D(\xi,t) = \rho_0 \int_{\mathbb{R}^3} \mathcal{Q}(X(\xi,t)-X(\zeta,t),t )\,d^3\zeta\,.
\end{equation}
The Lagrangian flow map is here given by the Zel'dovich ansatz, namely, 
\begin{equation}\label{Z13}
 X(\xi, t)= A(t)\, \big(\xi_1-f(\xi_1, t)\,, \xi_2\,, \xi_3\big)\,.
\end{equation}
Using Eq.~\eqref{3.12a}, the Lagrangian formula for the effective density of dark matter can be written as 
\begin{equation}\label{Z14}
 R_D = \frac{\rho_0\,\alpha_0\,\mu_0^3}{8\pi\, A(t)}\int_{\mathbb{R}^3}\frac{1+\chi}{\chi^2}\, e^{-\chi}\,d^3\zeta\,,
\end{equation}
where $\alpha_0=2/(\lambda_0\,\mu_0)\approx 11$ and 
 \begin{equation}\label{Z15}
 \chi:= \mu_0\, |X(\xi,t)-X(\zeta,t)|\,.
 \end{equation}
From the Zel'dovich ansatz for the Lagrangian flow map, we have
 \begin{equation}\label{Z16}
 X(\zeta,t)-X(\xi,t)=A(t)\,\Big(\zeta_1-\xi_1 + f(\xi_1,t) - f(\zeta_1,t)\,, \zeta_2-\xi_2\,, \zeta_3-\xi_3\Big)\,.
 \end{equation}
It is therefore useful to define a new integration variable $\sigma$ in Eq.~\eqref{Z14}, namely, 
 \begin{equation}\label{Z17}
 \zeta - \xi := \frac{\sigma}{\mu_0\, A(t)}\,.
 \end{equation}
Then, 
 \begin{equation}\label{Z18}
 \mu_0\,\left[ X(\zeta,t)-X(\xi,t) \right] = ({\cal S}\,, \sigma_2\,, \sigma_3)\,,
 \end{equation}
where 
 \begin{equation}\label{Z19}
{\cal S} := \sigma_1 + \mu_0\,A(t) \Big[f(\xi_1,t) - f(\xi_1+ \frac{\sigma_1}{\mu_0\,A(t)},t)\Big]\,.
 \end{equation} 
 
To work with dimensionless quantities, we define 
 \begin{equation}\label{Z20}
\gamma :=\mu_0\, \xi_1 \,, \qquad \tilde{\Sigma}(\gamma,t):= \mu_0\, f(\frac{\gamma}{\mu_0},t)\,,
 \end{equation}
so that 
 \begin{equation}\label{Z21}
{\cal S} = \sigma_1 + A(t) \Big[\tilde{\Sigma}(\gamma,t)-\tilde{\Sigma}(\gamma+\frac{\sigma_1}{ A(t)},t)\Big]\,.
 \end{equation} 
In this way, we find 
\begin{equation}\label{Z22}
 R_D = \frac{\rho_0\,\alpha_0}{8\pi\, A^4(t)}\int_{\mathbb{R}^3}\frac{1+\chi}{\chi^2}\, e^{-\chi}\,d^3\sigma\,,
\end{equation}
where 
 \begin{equation}\label{Z23}
 \chi = \sqrt{{\cal S}^2 + \sigma_2^2 + \sigma_3^2}\,
 \end{equation}
and ${\cal S}$ is given by Eq.~\eqref{Z21}.

It is convenient at this point to introduce spherical  polar coordinates 
\begin{equation}\label{Z24}
\sigma_1=\ell \cos\vartheta, \quad \sigma_2=\ell \sin\vartheta \cos\varphi\,, \quad   \sigma_3=\ell \sin\vartheta \sin\varphi\,
\end{equation}
with corresponding element of volume $\ell^2 \sin\vartheta\, d\ell \,d\vartheta \,d\varphi$. 
In these coordinates, $\chi$ given in Eq.~\eqref{Z23} does not depend on $\varphi$, since $\sigma_2^2+\sigma_3^2=\ell^2\sin^2\vartheta$. Using this fact, integration over $\varphi$ results in multiplication  of the remaining two-dimensional integral over $\ell$ and $\vartheta$ by $2\pi$.  With
 \begin{equation}\label{Z25}
\eta:=\cos \vartheta\,,
 \end{equation}
we have that
\begin{equation}\label{Z26}
 R_D = \frac{\rho_0\,\alpha_0}{A^4(t)}\,{\cal I}\,, \qquad {\cal I}= \frac{1}{4}\int_0^\infty\int_{-1}^1\,\frac{1+\chi}{\chi^2}\, e^{-\chi}\,\ell^2\, d\ell d\eta\,.
\end{equation}
Here,
 \begin{equation}\label{Z27}
 \chi = \sqrt{{\cal S}^2 + \ell^2\,(1-\eta^2)}\,
 \end{equation}
and
 \begin{equation}\label{Z28}
{\cal S} = \ell\,\eta + A(t) \Big[\tilde{\Sigma}(\gamma,t)-\tilde{\Sigma}(\gamma+\frac{\ell\,\eta}{ A(t)},t)\Big]\,.
 \end{equation} 
In terms of the new dimensionless quantities defined in Eq.~\eqref{Z20}, 
 \begin{equation}\label{Z29}
 \Psi = \frac{\partial\, \tilde{\Sigma}(\gamma,t)}{\partial\, \gamma} := \tilde{\Sigma}_{\gamma}(\gamma,t)\,
 \end{equation}
and the integro-differential Eq.~\eqref{Z9} of our model now takes the form
\begin{equation}\label{Z30}
\tilde{ \Sigma}_{\gamma tt}+2\,\frac{\dot{A}}{A}\, \tilde{ \Sigma}_{\gamma t} - \frac{4\pi G\rho_0}{A^3}\,(1+\frac{\alpha_0}{A})\,  
\tilde{ \Sigma}_{\gamma} = \mathcal{N}\,,
\end{equation}
where the nonlinear  part is given by
\begin{equation}\label{Z30a}
\mathcal{N}=\frac{4\pi G \rho_0\alpha_0}{A^4}\,\Big[(1-\tilde{\Sigma}_{\gamma})\mathcal{I}-1\Big]\,.
\end{equation}

Finally, it proves convenient to use $s=A(t)$ as the new temporal variable instead of the cosmic time $t$; that is, we define
\begin{equation}\label{Z31}
\Sigma(\gamma,s):=\tilde \Sigma(\gamma,A^{-1} (s))\,,
\end{equation}
so that 
\begin{align*}
\tilde\Sigma_{\gamma t}(\gamma,A^{-1}(s))&=\dot A(A^{-1}(s))\Sigma_{\gamma s}(\gamma,s)\,,\\
\tilde\Sigma_{\gamma tt}(\gamma,A^{-1}(s))&=\dot A^2(A^{-1}(s))\Sigma_{\gamma ss}(\gamma,s)+\ddot A(A^{-1}(s))\Sigma_{\gamma s}(\gamma,s)\,.
\end{align*}
We now use Eqs.~\eqref{cm:ode3}--\eqref{cm:ode4}, with $E=0$, 
to transform the integro-differential Eq.~\eqref{Z30} to  
\begin{align}
\label{Z32}
s^2(2 s+\alpha_0) \Sigma_{\gamma ss}+s(3 s+\alpha_0) \Sigma_{\gamma s}-3 (s+\alpha_0) \Sigma_{\gamma}&=
3\, \alpha_0 \,\Big[(1-\Sigma_{\gamma})\mathcal{I}(\Sigma)-1\Big]\,,
\end{align}
where $\mathcal{I}$ is the integral defined in display~\eqref{Z26} with $\chi$ given by Eq.~\eqref{Z27} and 
\begin{equation}\label{Z33}
\mathcal{S}= \ell\,\eta+s \Big[\Sigma(\gamma,s)-\Sigma(\gamma+\frac{\ell\,\eta}{ s},s) \Big]\,.
\end{equation}
Let us note that the dependence of $\mathcal{I}$ on $\Sigma$, through $\mathcal{S}$, is such that 
$\mathcal{I}(\Sigma)$ is unchanged if $\Sigma$ is replaced by $\Sigma + g(s)$, where $g(s)$ is an arbitrary function of $s$. Moreover, if $\Sigma$ is a solution of the integro-differential Eq.~\eqref{Z32}, then so is $\Sigma + g(s)$. In particular, it is simple to show that for $\Sigma=g(s)$
\begin{equation}\label{Z34}
\mathcal{I}(g)= 1\,, 
\end{equation}
so that 
\begin{equation}\label{Z35}
\Sigma (\gamma, s) = g(s)
\end{equation}
with $\Sigma_{\gamma}=0$ is an \emph{exact} solution of Eq.~\eqref{Z32}. 

The rest of this paper is about the nature of solutions of this nonlocal Zel'dovich model. 

\subsection{$\mathcal{N} \approx 0$}

To illustrate certain properties of our nonlocal  model, it proves useful at this point to define a limiting form of the model whose solution is in some sense the analog of the original Zel'dovich solution~\cite{Mu} in the present context. The Zel'dovich ansatz for our nonlocal toy model has led to Eq.~\eqref{Z32}, where, in contrast to the original Zel'dovich solution~\cite{Mu}, the nonlinear term that  appears on the right-hand side of this equation does not in general vanish. However, this nonlinear part of the  main differential equation vanishes for a certain limiting form of our toy model if we assume that 
\begin{equation}\label{4.1}
\rho_D(x, t) =\frac{\alpha_0}{A}\,  \rho(x, t)\,,
\end{equation}
where the scale factor $A(t)$ is given by Eq.~\eqref{cm:ode5} and $\rho(x, t)$ is the density of baryonic matter in the universe. In Eq.~\eqref{Z32}, the temporal variable $s$ is in fact $A(t)$, so that $s=0$ at the Big Bang and $s=1$ at the present epoch. The Zel'dovich ansatz in this limiting case ($\mathcal{N} = 0$) leads to the linear and homogeneous differential equation
\begin{equation}\label{4.2}
s^2(2 s+\alpha_0) \Sigma_{\gamma ss}+s(3 s+\alpha_0) \Sigma_{\gamma s}-3 (s+\alpha_0) \Sigma_{\gamma}=0\,,
\end{equation}
which is essentially the same as the linear perturbation equation~\eqref{3.10} for the density contrast $D_k(t)$ in the limit  where  $k\to 0$.

 With the change of variable $s=-\alpha_0\, \nu/2$ and some rearrangement, this differential equation for $\Sigma_\gamma (\gamma, s):= \Delta (\gamma, \nu)$ becomes
 \begin{align}\label{m:bt2}
\nu^2(1-\nu) \Delta_{\nu \nu}+\nu (1-\frac{3}{2}\,\nu)\Delta_{\nu}-3 (1-\frac{1}{2}\,\nu) \Delta=0. 
\end{align} 
The solution of this equation can be expressed as $\Delta(\gamma, \nu) = \mathcal{D}(\gamma)\, \mathcal{F}(\nu)$, where $\mathcal{D}(\gamma)$ is an arbitrary function of the spatial variable and $\mathcal{F}(\nu)$ satisfies Eq.~\eqref{m:bt2},  which can be solved in terms of the  hypergeometric function~\cite{A+S}. The end result is
 \begin{equation}\label{hypans}
\Sigma_\gamma (\gamma, s) = \mathcal{D}(\gamma)\, S_p(s)\,, 
 \end{equation}
where
 \begin{equation}\label{hypans'}
 S_p(s)= s^p F(p+\frac{3}{2},p-1, 2 p+1; -\frac{2s}{\alpha_0})\,.
 \end{equation}
Here, $p=\pm\sqrt{3}$,  $F$ is the hypergeometric function~\cite{A+S}, $\nu=-2s/\alpha_0$ and $|\nu|<1$.  We note that $S_{\sqrt{3}}$ and  $S_{-\sqrt{3}}$  form a fundamental set of solutions of the linear second-order differential Eq.~\eqref{m:bt2}. The general solution of  Eq.~\eqref{m:bt2} is qualitatively similar to the classical Zel'dovich solution; in particular,  the given fundamental set of solutions consists of one growing and one decaying  mode.

\begin{figure}
\centerline{\includegraphics[width=3in]{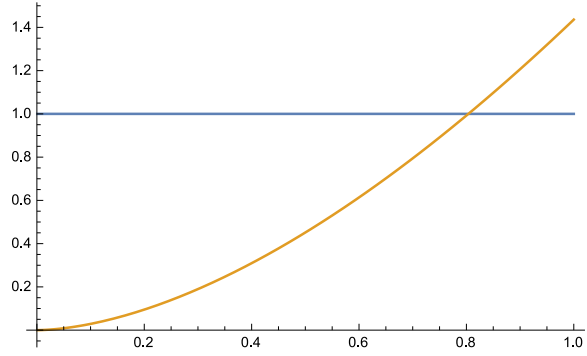}}
\caption{Plot of $s\mapsto 10^{-5} S_{\sqrt{3}}(s)  / S_{\sqrt{3}}(10^{-3})$.\label{fig: fig3} }
\end{figure}

Under what conditions would $\rho_D(x, t)$, given by Eq.~\eqref{dark}, reduce to Eq.~\eqref{4.1}? That is, we wish to determine the conditions under which the limiting form of the model under consideration here approximates our basic nonlocal model. Let us note in this connection that kernel $\mathcal{Q}$ in Eq.~\eqref{3.12a} is radial and decays exponentially for $|r| \gg \mu_0^{-1}$.  Therefore, if $\rho (x, t)$ always  varies in space only over scales much larger than $\mu_0^{-1}$, then its Fourier integral transform is $\hat \rho(k, t)$,  
\begin{equation}\label{4.2.1}
\hat \rho(k, t)= \int \rho (r, t)\, e^{-ik \cdot r/A(t)}\, d^3r\,,
\end{equation}
which is essentially confined to a region in Fourier space such
 that $|k| \ll \mu_0 A(t)$. For instance, if we imagine that $\rho(x, t)$ always has the form of a Gaussian distribution centered around $x=0$ with a root-mean-square deviation from the mean of $\sigma_0 \gg \mu_0^{-1}$, then in the Fourier domain, this corresponds to a Gaussian distribution centered around $k=0$ with a root-mean-square deviation from the mean of $\hat \sigma_0 \ll \mu_0\, A(t)$. We can use this fact in conjunction with  the convolution theorem for Fourier transforms to show that the solution of the nonlocal model approaches the exact solution of the $\mathcal{N} = 0$ model for large  scale deviations from spatial homogeneity. That is, Eq.~\eqref{dark} can be written in the Fourier domain, in the sense defined in section IV, as 
\begin{equation}\label{4.2.2}
\hat \rho_D(k, t)= Q_k(t)\, \hat \rho(k, t)\,,
\end{equation}
where $\hat \rho_D(k, t)$ is the Fourier integral transform of $\rho_D$ and $Q_k(t)$ is given by Eq.~\eqref{3.12}. It is clear from Figure 1 that if $|k| \ll \mu_0 A(t)$, then $Q_k(t) \approx \alpha(t)$, where $\alpha(t)=\alpha_0/A(t)$, so that 
\begin{equation}\label{4.2.3}
\hat \rho_D(k, t) \approx \alpha(t)\, \hat \rho(k, t)\,.
\end{equation}
The inverse Fourier integral transform of this relation amounts essentially to Eq.~\eqref{4.1}. Therefore, on large scales that persist over time, the solution of the nonlocal model corresponding to the growing mode should approach
$S_{\sqrt{3}}(s)$. In Figure 3, we plot 
\begin{equation}\label{4.2.4}
h (s)= 10^{-5}\,\frac{S_{\sqrt{3}}(s)}{S_{\sqrt{3}}(10^{-3})}\,,
\end{equation} 
where it is assumed that at the epoch of recombination corresponding to cosmological redshift  $z \sim 10^{3}$ and  $A=1/(1+z)$, 
the density contrast is $\sim 10^{-5}$. It is demonstrated in Figure 3 that $h$ approaches unity well  before the present era $s=1$. 

Let us recall here that as $\Sigma_\gamma \to 1$, the density of baryons $\rho_0/[A^3\,(1-\Sigma_\gamma)]$ approaches infinity. At some stage in this process, the density contrast is so large that the perturbation separates from the background and collapses under its own gravitational attraction, thus leading to the formation of structure in the universe~\cite{Pe, ZN, Mu}. The result presented in Figure 3 indicates that the nonlocal model under consideration here is such that for  sufficiently large scale perturbations after recombination, structure formation is theoretically possible in this toy model.

\section{An Exact Solution}

Our nonlocal model has an exact nontrivial solution. To this end, let 
\begin{equation}\label{7.1}
\Sigma (\gamma, s) = g(s) + \beta(s) \,\gamma\,,
\end{equation}
where $g$ and $\beta$ are only  functions of the temporal variable $s$. It proves interesting to start by computing  $\mathcal{I}(\Sigma)(\gamma,s)$ given in Eq.~\eqref{Z26}. We find that  $\mathcal{S}=\ell \,\eta\,(1-\beta)$ and $\chi=\ell \, \chi_0$, where $\chi_0$ is independent of  $\ell$ and is given by
\begin{equation}\label{7.2}
\chi_0 := [\,\eta^2\,(1-\beta)^2+1-\eta^2\,]^{1/2}\,.
\end{equation}
The integration in $\mathcal{I}$ over $\ell : 0 \to \infty$ can be carried out first using the relation
\begin{equation}\label{7.3}
 \int_0^{\infty} \ell^{\,n}\,e^{-\ell\, \chi_0}\, d\ell = \frac{n!}{\chi_0^{n+1}}\,
\end{equation}
for $n= 0, 1, 2, \ldots$. For the integration over $\eta : -1 \to 1$, we note that 
\begin{equation}\label{7.3A}
\int \frac{x^{2n}}{(1+c\,x^2)^{n+3/2}}\,dx= \frac{1}{2n+1}\, \frac{x^{2n+1}}{(1+c\,x^2)^{n+1/2}}\,.
\end{equation}
Therefore, for $c>-1$,
\begin{equation}\label{7.3B}
\int_{-1}^1 \frac{\eta^{2n}}{(1+c\,\eta^2)^{n+3/2} }\,d\eta = \frac{2}{2n+1}\, \frac{1}{(1+c)^{n+1/2} }\,,
\end{equation}
for $n=0, 1, 2, \ldots$. For $n=0$, we find that
\begin{equation}\label{7.4}
\mathcal{I}(\Sigma) = \int_0^1\frac{d\eta}{\{1+[(1-\beta)^2-1] \, \eta^2\}^{3/2}}=\frac{1}{|1-\beta|}\,.
\end{equation}
It follows from this result that for $\beta<1$, the right-hand side of Eq.~\eqref{Z26} vanishes, so that 
$\Sigma_\gamma= \beta(s)$ must satisfy
\begin{equation}\label{7.5}
s^2(2 s+\alpha_0) \beta_{ss}+s(3 s+\alpha_0) \beta_{s}-3 (s+\alpha_0) \beta=0\,.
\end{equation}
Thus $\beta$ can be written as 
\begin{equation}\label{7.6}
\beta (s)=C_{+}\, S_{\sqrt{3}}(s) + C_{-}\, S_{-\sqrt{3}}(s)\,,
\end{equation}
where $S_{\pm\,\sqrt{3}}(s)$ are given in Eq.~\eqref{hypans'} and  $C_{\pm}$ are constants such that $\beta(s)<1$. 

Let $\tilde{B}(t) :=\beta (A(t))$, then the exact solution under consideration corresponds to the Zel'dovich ansatz~\eqref{Z1} such that
\begin{equation}\label{7.7}
X_1=A(t)\, [1-\tilde{B}(t)]\, \xi_1\,, \qquad X_2=A(t)\, \xi_2\,, \qquad X_3=A(t)\, \xi_3\,,
\end{equation}
which represents an expanding homogeneous but \emph{anisotropic} cosmological model. It is straightforward to check that with the spatially uniform matter density $\rho_0/[A^3\,(1-\tilde{B})]$, our original model Eqs.~\eqref{cm:prepeqsa}--\eqref{cm:prepeqsb} are satisfied in this case.

If $C_{+}>0$ in Eq.~\eqref{7.6}, then $\beta$ will increase monotonically with respect to increasing time and eventually approach unity, in which case the oblate spheroidal model universe collapses to form an expanding circular disk of matter. 

It was argued at the end of  section V that the solution of the nonlocal model should approach the solution of the $\mathcal{N}=0$ model if the density $\rho(x, t)$ always  varies in space over distances that are much larger than $\mu_0^{-1}$. The limiting situation, where $\rho(x, t)$ loses all dependence upon space and depends only upon time corresponds to 
\begin{equation}\label{7.8}
\Sigma_\gamma (\gamma, s)=\beta(s)\,,
\end{equation}
from which the exact solution under consideration would necessarily follow. These remarks then provide the physical interpretation for the exact solution presented here.

\subsection{Linear Perturbation of the Exact Solution}

Consider the linear operator
\begin{equation}\label{7.9}
\mathcal{L} := s^2(2 s+\alpha_0)\, \frac{\partial^2}{\partial s^2}+s(3 s+\alpha_0)\, \frac{\partial}{\partial s} -3 (s+\alpha_0)\,;
\end{equation}
then,  our integro-differential Eq.~\eqref{Z32} can be written as 
\begin{equation}\label{7.10}
\mathcal{L}\, \Sigma_\gamma =3\, \alpha_0 \,\Big[(1-\Sigma_{\gamma})\mathcal{I}-1\Big]\,.
\end{equation}
This equation has an exact solution 
\begin{equation}\label{7.11}
\Sigma^0 (\gamma, s) = g(s) + \beta(s) \,\gamma\,,
\end{equation}
where $g(s)$ is arbitrary and $\beta(s) <1$ is a solution of the homogeneous linear equation $\mathcal{L} \,\beta(s)=0$. We would now like to look for a solution of Eq.~\eqref{7.10} of the form
\begin{equation}\label{7.12}
\Sigma = \Sigma^0 +\epsilon\, \Pi (\gamma, s)\,,
\end{equation}
where $\Pi$ is the perturbing function and $\epsilon$, $0<\epsilon \ll 1$, is the perturbation parameter such that only terms linear in $\epsilon$ will be considered. To compute $\mathcal{I}(\Sigma)(\gamma,s)$ in this case, we note that 
\begin{equation}\label{7.13}
\mathcal{S}=\ell \,\eta\,(1-\beta) - \epsilon\, \mathcal{T}\,,
\end{equation}
where 
\begin{equation}\label{7.14}
\mathcal{T} := s\,\left[\Pi(\gamma+\frac{\ell\,\eta}{s}, s)-\Pi(\gamma,s)\right]\,.
\end{equation}
Hence, 
\begin{equation}\label{7.15}
\chi= \ell \, \chi_0 -\epsilon\, \eta\,(1-\beta)\, \chi_o^{-1}\,\mathcal{T}\,
\end{equation}
and 
\begin{equation}\label{7.16}
 \frac{1+\chi}{\chi^2}\, e^{-\chi} =  \frac{1+\ell \, \chi_0}{\ell^2\, \chi_0^2}\, e^{-\ell \, \chi_0}+\epsilon\, \eta\,(1-\beta)\,\mathcal{T}~ \frac{2+2\,\ell \, \chi_0 +\ell^2 \, \chi_0^2}{\ell^3\, \chi_0^4}\, e^{-\ell \, \chi_0}\,.
\end{equation}

To proceed, assume that perturbation $\Pi$ is analytic in spatial variable $\gamma$ so that $\mathcal{T}$ is represented by its Taylor series,
\begin{equation}\label{7.17}
\mathcal{T} = \sum_{m=1}^{\infty} \frac{1}{m!}\, \Pi^{(m)}(\gamma, s) \, \frac{\ell^m\, \eta^m}{s^{m-1}}\,,
\end{equation}
where $\Pi^{(m)} := \partial^{\,m}\,\Pi/\partial \gamma^{\,m}$. Moreover, we can write 
\begin{equation}\label{7.18}
\mathcal{I}(\Sigma) = \mathcal{I}(\Sigma^0) + \epsilon\, \mathcal{K}\,,
\end{equation}
where for $\beta <1$, we have from Eq.~\eqref{7.4} that 
\begin{equation}\label{7.19}
\mathcal{I}(\Sigma^0) =\frac{1}{1-\beta}\,.
\end{equation}
In evaluating $\mathcal{K}$, the integration over $\ell$ is straightforward using Eq.~\eqref{7.3} and we find
\begin{equation}\label{7.20}
 \int_0^{\infty} \ell^{\,m-1}\,[\ell^2\, \chi_0^2 + 2\, \ell\, \chi_0 +2]\, e^{-\ell\, \chi_0}\, d\ell = \frac{(m+1)(m+2)(m-1)!}{\chi_0^{m}}\,.
\end{equation}
The subsequent integration over $\eta$ vanishes by symmetry unless the integrand is even in $\eta$, which means that $m$ must be odd, namely, $m=2n+1$, where $n=0,1, 2, \ldots$. Then, using Eq.~\eqref{7.3B}, we find 
\begin{equation}\label{7.21}
\mathcal{K} = \sum_{n=0}^{\infty} \frac{n+1}{2n+1}\, \frac{\Pi^{(2n+1)}(\gamma, s)}{s^{2n}\,(1-\beta)^{2n+2}}\,.
\end{equation}
It follows from Eq.~\eqref{7.10}  that to first order in $\epsilon$, we have the equation for the  linear perturbation of the exact solution, namely, 
\begin{equation}\label{7.22}
\mathcal{L}\, \Pi_\gamma(\gamma, s) =3\, \alpha_0 \,\Big[(1-\beta)\mathcal{K}-\frac{\Pi_\gamma}{1-\beta}\Big]\,.
\end{equation}
Combining Eqs.~\eqref{7.21} and~\eqref{7.22}, we finally get
\begin{equation}\label{7.23}
\mathcal{L}\, \Pi_\gamma(\gamma, s) =3\, \alpha_0 \sum_{n=1}^{\infty} \frac{n+1}{2n+1}\, \frac{\Pi^{(2n+1)}(\gamma, s)}{s^{2n}\,(1-\beta)^{2n+1}}\,.
\end{equation}

\subsection{Solutions of the Perturbation Equation}

Inspection of Eq.~\eqref{7.23} for the linear perturbation $\Pi$ reveals that 
\begin{equation}\label{7.24}
\Pi= \frac{1}{2}\,\theta(s)\, \gamma^2\,, \qquad   \mathcal{L}\, \theta(s)=0\,
\end{equation}
is a solution, since the right-hand side of Eq.~\eqref{7.23} vanishes identically in this case.

Another possibility involves a solution of Eq.~\eqref{7.23} of the form
\begin{equation}\label{7.25}
\Pi= e^{b\,\gamma}\, \mathcal{E}(s)\,,
\end{equation}
where $b$ is a constant. Substituting Eq.~\eqref{7.25} into Eq.~\eqref{7.23} results in 
\begin{equation}\label{7.26}
\mathcal{L}\, \mathcal{E}(s) =\frac{3\, \alpha_0}{b}\, s\, \mathcal{E}(s)\,\sum_{n=1}^{\infty} \frac{n+1}{2n+1}\, \mathbb{B}^{2n+1}\,,
\end{equation}
where
\begin{equation}\label{7.27}
\mathbb{B}(s) :=\frac{b}{s\,(1-\beta)}\,.
\end{equation}
We recall that $s$ is the scale factor and for an expanding universe model, $s : 0 \to \infty$. Moreover, $\beta(s) < 1$. If we assume that the constant $|\,b\,|$ is such that $|\mathbb{B}(s)| < 1$, then the series in Eq.~\eqref{7.26} converges uniformly and we have 
\begin{equation}\label{7.28}
\sum_{n=1}^{\infty} \frac{n+1}{2n+1}\, \mathbb{B}^{2n+1}=\frac{1}{2} \Big[\frac{\mathbb{B}^3}{1-\mathbb{B}^2} - \mathbb{B} - \frac{1}{2} \ln{\left(\frac{1-\mathbb{B}}{1+\mathbb{B}}\right)}\Big] := \mathcal{B}(s)\,.
\end{equation}
The function $\mathcal{E}(s)$ can now be determined from the linear differential equation
\begin{equation}\label{7.29}
\mathcal{L}\, \mathcal{E}(s) =\frac{3\, \alpha_0}{b}\, s\, \mathcal{B}(s)\,\mathcal{E}(s)\,.
\end{equation}

These possible solutions of the linear perturbation equation  indicate that the exact solution of our nonlocal model is \emph{unstable}. A complete analysis of the solutions of Eq.~\eqref{7.23} is beyond the scope of this work. 

\subsection{$\Sigma^0 = g(s)$}

Finally, it is important to point out a seemingly trivial special case that will be shown to be of physical significance in the next section, namely, when $\beta(s)=0$. In this case, $\Sigma^0_\gamma=0$ and we start with a small density perturbation of $\epsilon\, \Pi_\gamma$. The linear perturbation away from this exact zero solution can be unstable due to solution~\eqref{7.24} leading to $\Pi_\gamma=\theta(s)\,\gamma$ as well as solution~\eqref{7.25}, which diverges exponentially in the spatial variable $\gamma$, where $b$, $|\,b\,| <s$, can be positive or negative.

\section{Solution of the Nonlocal Model}

In our nonlocal Zel'dovich model, the density of baryons is given by $\rho_0/[s^3(1-\Sigma_\gamma)]$, where $s=A(t)$ is the scale factor and $\Sigma$ is a solution of our integro-differential Eq.~\eqref{Z32}. In the simplest situation of physical interest, we would like to start out with, say, a Gaussian density perturbation after the recombination epoch at $s=s_1$, namely, 
\begin{equation}\label{S1}
 \Sigma_\gamma(\gamma, s_1)= \delta\,\frac{2}{\sqrt{\pi}}\, e^{-(\gamma/\Gamma)^2}\,,
\end{equation}
where $\delta$ is a positive constant that denotes the strength of the small perturbation away from a homogeneous distribution of baryonic matter and 
$\Gamma$  is a spatial scaling constant related to the spatial width of the Gaussian perturbation. Here $\Sigma (\gamma, s_1)$ can be determined from a simple spatial integration. We recall that if $\Sigma$ is a solution of Eq.~\eqref{Z32}, then so is $\Sigma + g(s)$;  now,  taking advantage  of this arbitrariness in $\Sigma$, we  set $\Sigma(0, s_1)=0$ with no loss in generality. Thus integrating Eq.~\eqref{S1} over $\gamma$ with this boundary condition, we obtain
\begin{equation}\label{S2}
 \Sigma(\gamma, s_1)= \delta\, \Gamma\,\frac{2}{\sqrt{\pi}}\, \int_0^{\,\gamma/\Gamma} e^{-x^2} \,dx\,.
\end{equation}
Let us recall here  the definition of the \emph{error function}~\cite{A+S},
\begin{equation}\label{S3}
{\rm erf} (z) =\frac{2}{\sqrt{\pi}}\, \int_0^z e^{-x^2} \,dx\,,
\end{equation} 
where ${\rm erf}(\infty)=1$. Thus the initial perturbation in $\Sigma$ is an odd function of $\gamma$ such that $ \Sigma(\gamma, s_1) : -\delta\, \Gamma \to \delta\, \Gamma$  for $\gamma : -\infty \to \infty$. We would like to solve the integro-differential Eq.~\eqref{Z32} to determine how such a perturbation would evolve in time for $s_1 \approx 10^{-3}$ and $\delta \approx 10^{-5}$. No such exact solution is known; therefore, we must proceed numerically. However, we know from the linear perturbation approach that for sufficiently small $\delta$, the linear perturbation away from the exact zero solution is unstable; hence,  the numerical solution in this case would be fraught with instabilities.

\subsection{Numerical Approach}

To prepare for the numerical work, we reformulate partial integro-differential Eq.~\eqref{Z32}  as  a  system of first-order nonautonomous  ordinary differential equations (ODEs) in an infinite-dimensional function space. It proves advantageous for the numerical work to introduce a new temporal variable given by $\tau$, related to the scale factor via  $s = A(t) = \exp{(\tau)}$. Therefore, we start with partial  integro-differential Eq.~\eqref{Z32} and let $s=e^\tau$. Then,
\begin{equation}\label{5.1}
s\,\Sigma_{\gamma\,s}=  \Sigma_{\gamma\,\tau}\,, \qquad s^2\,\Sigma_{\gamma\,s\,s}=  \Sigma_{\gamma\,\tau\, \tau} - \Sigma_{\gamma\,\tau}\,.
\end{equation}
Hence, Eq.~\eqref{Z32} can be written as 
\begin{equation}\label{5.2}
(2 e^\tau + \alpha_0)  \Sigma_{\gamma\,\tau\, \tau} + e^\tau  \Sigma_{\gamma\,\tau}-3 ( e^\tau + \alpha_0)  \Sigma_{\gamma}=3\,\alpha_0 \big[(1-\Sigma_{\gamma})\mathcal{I}-1\big]\,.
\end{equation}

Next, let $\Sigma =\Sigma (\gamma, \tau)$ and define $\Phi$ and $\Upsilon$ such that 
 \[\Phi(\gamma,\tau):=\Sigma_{\gamma}(\gamma,\tau)\,,\qquad \Upsilon:=\Phi_\tau\,,\]
 so that, as before, 
 \[\Sigma(\gamma, \tau) =\int_0^{\,\gamma}\Phi(x, \tau)\,dx\]
 and
 \[ \Sigma_\tau (\gamma, \tau) =\int_0^{\,\gamma}\Upsilon(x, \tau)\,dx.\]
Using these formulas, we finally arrive at the system that we numerically integrate, namely,  
 \begin{align} \label{5.3} 
\nonumber  \Sigma_\tau&=\int_0^{\,\gamma} \Upsilon (x, \tau)\,dx\,,\\
\nonumber \Phi_\tau &= \Upsilon\,,\\
(2 e^\tau+\alpha_0) \Upsilon_\tau &= - e^\tau \Upsilon+3 (e^\tau+\alpha_0) \Phi+N\,,
\end{align}
where the nonlinear term $N$ is given by
\begin{equation}\label{5.3a}
N = 3\,\alpha_0 \Big[(1-\Phi)\mathcal{I}(\Sigma)-1\Big]\,.
\end{equation}
We integrate this system from  $\tau=\tau_1$ to $\tau =0$.
For $N=0$, the solution of this system is equivalent to the linear solution described in part  B of section V.

It is remarkable that  solving for $\mathbb{F}$ in the Zel'dovich ansatz has been reduced to solving a nonautonomous first-order system of ODEs  in some function space (of functions of $\gamma$). While this first-order system is singular at $\tau=-\infty$ (which according to the definition of $A$ is the instant of the Big Bang), at the decoupling epoch $\tau=\tau_1\sim -3\, \ln 10$ and beyond, the system is nonsingular at least so long as the Lagrangian flow map is defined.  In the present notation, the  physically relevant solution is defined as long as $\Phi<1$.  At $\Phi=1$ the density of matter becomes infinite as previously explained.

System~\eqref{5.3} has a rest point $(\Sigma,\Phi,\Upsilon) =0$, which is the exact zero solution of our nonlocal model discussed before.  It has been shown in section VI  that a linear perturbation away from this point is unstable.  Moreover,  system~\eqref{5.3} is expected to have a unique solution for a state $(\Sigma,\Phi,\Upsilon)$  given at  $\tau=\tau_1$  as long as the initial functions (of $\gamma$) belong to a Banach space of smooth functions.  The analysis of this problem is beyond the scope of the present paper. 

In our numerical work, the initial perturbation is naturally taken to be a Gaussian. We choose the initial conditions for the integration of system~\eqref{5.3} such that for $N=0$ the \emph{growing mode} of the linear solution discussed in part B of section V is recovered. Specifically,  using the error function and the new function $\omega$ defined by
\[
\omega(\tau):=e^{[(\tau-\tau_1)\,\sqrt{3}]}\, \frac{F(\sqrt{3}+\frac{3}{2}, \sqrt{3}-1,2\sqrt{3}+1;-\frac{2 e^\tau}{11})}{F(\sqrt{3}+\frac{3}{2}, \sqrt{3}-1,2\sqrt{3}+1;-\frac{2 e^{\tau_1}}{11})}\,,
\]
where $F$ is the hypergeometric function as in  display~\eqref {hypans'}, the initial data at $\tau=\tau_1$ are  taken to be
\begin{equation}\label{5.4}
 \Sigma(\gamma,\tau_1)= \delta\, \Gamma\frac{2}{\sqrt{\pi}}\, \int_0^{\,\gamma/\Gamma} e^{-x^2} \,dx\,,
\end{equation}
\begin{equation}\label{5.5}
 \Phi(\gamma,\tau_1)= \delta\,\frac{2}{\sqrt{\pi}}\, e^{-(\gamma/\Gamma)^2}\,,
\end{equation}
\begin{equation}\label{5.6}
 \Upsilon(\gamma,\tau_1)= \delta\,\frac{2}{\sqrt{\pi}}\, e^{-(\gamma/\Gamma)^2}\,\omega'(\tau_1)\,.
\end{equation}
Here,  $\omega'(\tau) :=d\omega/d\tau$  and $\Gamma= 10, 10^2, 10^3$, and so on, is the spatial scale factor  such that $\Gamma \, \mu_0^{-1}$ is proportional to the width of the Gaussian perturbation.

\subsection{Numerical Algorithm}

Determining the behavior of solutions of  the nonlinear, non-autonomous infinite-dimensional system of ODEs~\eqref{5.3}  with initial data~\eqref{5.4}--\eqref{5.6} seems to be beyond current understanding of infinite-dimensional ODEs. For this reason, we approach the problem via numerical experiments. It is important to note here that, due to the nature of the subject matter, some of the standard notation employed here is independent of the rest of this paper.

The state $(\Sigma,\Phi,\Upsilon)$ in system~\eqref{5.3} is a vector of functions of the real variable $\gamma$ on the whole real line. These states evolve with respect to the temporal variable $\tau$ from the initial state at $\tau=\tau_1$. Fortunately, the initial data represents a localized change (perturbation) of the baryonic matter in the background flow so that, up to a close approximation,  it  vanishes outside of some finite interval  containing the origin of our (Lagrangian) spatial coordinate $\gamma$. While discretization in general would require restricting the domain of the evolving state to  some finite interval, this restriction is natural here due to the localized initial data.  Thus, we choose an appropriate  number $L>0$ and set the corresponding  computational domain to be the interval $-L\le \gamma \le L$.  
To discretize, we fix an even positive integer $m$ and define the spatial increment 
\[\Delta \gamma=\frac{2L}{m}\] 
so that the discrete space variable is a vector of length $m+1$ with components
\[\gamma_i: = -L+ (i-1) \Delta\gamma , \qquad i=1,2,3,\ldots,m+1.\]
In view of this choice of spatial discretization,  function $\Sigma$ is discretized  in the usual manner as well,
\[ \Sigma_i=\Sigma(\gamma_i,\tau)\,,\]
and the same scheme is used for $\Phi$ and $\Upsilon$. 

To complete the spatial discretization of the time dependent vector field that defines the infinite-dimensional ODE requires spatial discretizations of the integrals that appear on the right-hand sides of the equations in system~\eqref{5.3}. 
Integration is discretized using a combination of the trapezoidal rule, the composite Simpson's $1/3$ rule  and Simpson's $3/8$ rule (see Ref.~\cite{B+F},  p.~195).
For the integral in the first component of the infinite-dimensional vector field, evaluation is required for each $\gamma_i$ in the interval $[-L,L]$. The discrete value of the integral  for  data $\Upsilon_i$, which  represent the discretization of the function $\gamma \mapsto \Upsilon(\gamma,\tau)$ at a specified value of $\tau$, is approximated according to the position of the $i$th  node in the interval $[-L,L]$. For $\gamma_i$ at node $m/2+1$ (corresponding to $\gamma=0$) the value of the integral is of course zero. For $i>m/2+1$ there are several cases: The trapezoidal rule is used when $i=m/2+2$; Simpson's composite $1/3$ rule is used if the number of nodes from $m/2+1$ to $i$ inclusive is odd (so that the number of subintervals is even),  Simpson's $3/8$ rule is used on the interval corresponding to the first four nodes and Simpson's composite $1/3$ rule is used on the remaining interval corresponding to the remaining odd number of nodes in case the number of nodes from $m/2+1$ to $i$ inclusive  is even. A similar scheme  is used in case $i< m/2+1$ (with attention to the sign change in the integration when $\gamma<0$).  This discretization converts the first equation in system~\eqref{5.3} into $m+1$ ordinary differential equations in the $3m+3$  variables $(\Sigma_i,\Phi_i,\Upsilon_i)$,  where $i=1,2,3,\ldots,m+1$.
The obvious discretization of the second equation in system~\eqref{5.3} produces $m+1$ equations in the $3m+3$ (discrete) state variables. 
The third equation of the system also produces  $m+1$ ODEs in the $3m+3$ unknowns  after a suitable discretization of the  operator $\mathcal {I}$ given by
\begin{align}\label{bri}
\mathcal{I}(\Sigma)(\gamma,\tau)&=\frac{1}{4}\int_0^\infty\int_{-1}^1 \frac{1 +\sqrt{\mathcal{S}^2+\ell^2(1-\eta^2)}}{\mathcal{S}^2+\ell^2(1-\eta^2)}\,e^{-\sqrt{\mathcal{S}^2+\ell^2(1-\eta^2)}}\,\ell^2\, d\eta\, d\ell,
\end{align}
where
\[
\mathcal{S:}=\ell \eta+e^\tau \big[\Sigma(\gamma,\tau)-\Sigma(\gamma+\frac{\ell\,\eta}{e^\tau},\tau)\big].
\]
One ODE is produced for each index $i$, $i=1,2,3,\ldots,m+1$.
While numerical approximation of double integrals is problematic in general, approximation of the  double integral in display~\eqref{bri} requires a careful treatment for at least three reasons:  integration over $0\le \ell<\infty $ must be restricted to a finite interval,  the denominator of the integrand vanishes at $\ell=0$,  and it also vanishes at $\eta=\pm 1$ whenever 
\begin{equation}\label{apzexp}
 \pm \ell+e^\tau \big[\Sigma(\gamma,\tau)-\Sigma(\gamma+\frac{\pm \ell}{e^\tau},\tau)\big]=0\,.
 \end{equation}

A key observation is that the denominator does not vanish at $\eta=\pm 1$ for $\ell>0$; that is, the third possibility does not occur over the domain of integration. To check this, simply note that the left-hand side of  Eq.~\eqref{apzexp} vanishes at $\ell=0$ and its derivative with respect to $\ell$ is 
\[\pm[1-\Phi(\gamma+\frac{\pm\ell}{e^\tau},\tau)]\,.
\]
The latter expression is positive for the plus sign and negative for the minus sign because, as previously stated, the physically relevant solution of the dynamical model requires that  $\Phi<1$. In particular,  expression~\eqref{apzexp} does not vanish for $\ell>0$ and the integrand  of the double integral is not singular at $\eta=\pm 1$ except at $\ell=0$. At these values of $\eta$ and $\ell$, the denominator vanishes. 

The singularities at $\eta=\pm 1$ and $\ell=0$ are removable. Indeed, by a Taylor series expansion of $\mathcal{S}$ with respect to $\ell$ at $\ell=0$, we find
\begin{equation}\label{apzexpA}
\mathcal{S}= [1-\Phi(\gamma,\tau)]\,\eta\,\ell+O(\ell^2)\,.
\end{equation}
It follows that the leading term of the corresponding expansion of $\mathcal{S}^2$ is of second order in $\ell$. Thus, the singularity is removable because  the  numerator of the integrand contains the factor $\ell^2$.

To utilize this analysis in the numerical algorithm, a  real number $\ell_0$,  $0<\ell_0\ll 1$,  is chosen  and the factor
\[\frac{\ell^2}{\mathcal{S}^2+\ell^2(1-\eta^2)}\] of the integrand is  approximated,   whenever $0\le \ell<\ell_0$,  by
\[
\frac{1}{[1-\Phi(\gamma,\tau)]^2\eta^2+1-\eta^2}\,.
\]
In effect, the contribution from the interval $\ell \in (0, \ell_0)$ to the integral $\mathcal{I}$ in Eq.~\eqref{bri} is thus given by
\begin{equation}\label{A}
\ell_0\,\int_{-1}^{1} \frac{d\eta}{[1-\Phi(\gamma,\tau)]^2\eta^2+1-\eta^2}= \frac{\ell_0}{\Omega}\,\ln{\Big(\frac{1+\Omega}{1-\Omega}\Big)}\,,
\end{equation}
where
\begin{equation}\label{B}
\Omega = \sqrt{1-(1-\Phi)^2}\,.
\end{equation}

Next, for  the contribution from the remaining interval $\ell \in (\ell_0, \infty)$ to the integral $\mathcal{I}$ in Eq.~\eqref{bri}, we substitute instead the integration over an appropriately chosen  finite interval 
$(\ell_0, K)$  for some sufficiently large $K>0$, which is justified by the rapid decay  (with respect to increasing $\ell$) of the negative exponential factor  in the integrand. Then, positive even integers $k$ and $n$ are selected and increments $\Delta \ell$ and $\Delta \eta$ are defined as
\begin{equation}\label{C}
\Delta\ell:=\frac{K-\ell_0}{k}\,,\qquad \Delta\eta:=\frac{ 2}{n}\,,
\end{equation}
so that the integration domain is covered by the natural rectangular grid given by
\begin{equation}\label{D}
\ell_{\hat{i}}:= \ell_0 + (\hat{i} -1) \Delta\ell, \qquad  \eta_j=-1+(j-1) \Delta\eta\,,
\end{equation}
where $\hat{i}=1,2,3,\ldots,k+1$ and $j=1,2,3,\ldots,n+1$. To proceed, the following
 inner integral is used to define the function  $I$ of $\ell$, $\gamma$,  $\tau$  and $\Sigma$ given by
\begin{equation}\label{E}
I(\ell,\gamma,\tau,\Sigma)=\int_{-1}^{1} \frac{1 +\sqrt{\mathcal{S}^2+\ell^2(1-\eta^2)}}{\mathcal{S}^2+\ell^2(1-\eta^2)}\,e^{-\sqrt{\mathcal{S}^2+\ell^2(1-\eta^2)}}\,\ell^2\, d\eta
\end{equation}
and  $I(\ell_i,\gamma, \tau,\Sigma)$ is  approximated by the composite Simpson's $1/3$ rule over $\eta$ for fixed $\gamma$,  $\tau$ and $\Sigma$. Of course, in this scheme, the outer integral 
is viewed as 
\begin{equation}
\int_{\ell_0} ^K I(\ell,\gamma, \tau,\Sigma)\,d\ell
\end{equation}
and again it is approximated for fixed $\gamma$,  $\tau$, and $\Sigma$ using the composite Simpson's $1/3$ rule with respect to integration over $\ell$. Thus, with obvious discretization of its local part,  the third equation of the infinite-dimensional system is approximated by $m+1$ ODEs. 
Initial data for the infinite-dimensional system is discretized in the same manner as the state variables over the grid points $\gamma_i$, $i=1,2,3,\ldots,m+1$. 
While the resulting initial value problem for the $(3m+3)$-dimensional system of ODEs may be passed to a general-purpose ODE solver to approximate its solution,  sufficiently fine spatial resolution and the computational expense incurred by the double integration required at each function evaluation suggests employment of an efficient time stepping routine with few function evaluations per step. A natural choice is the midpoint method (also called the leapfrog method). This method was implemented with restart (after each 20 time steps) to mitigate its well-known numerical instability.  In addition, each block of 20 steps was recomputed with half the original step size followed by one Richardson extrapolation. Due to the difference in the computation using Simpson's $1/3$ rule and Simpson's $3/8$ rule for odd and even numbers of grid points associated with the discretized variable $\gamma$  for the  numerical approximation of the integral in the first component of the infinite-dimensional vector field,   the discretized functions  $\Sigma$, $\Phi$ and $\Upsilon$ incur small numerical oscillations. These were mollified by filtering  with  two-point averaging before starting the next block of 20 steps. The time stepping used here, which requires only two function evaluations per step,  is fourth-order accurate in the increment $d\tau$. Simpson's rules  for numerical integration are both fourth-order accurate. Thus the entire scheme is fourth-order accurate in the space and time increments except for the trapezoidal rule numerical integrations when function values at only two grid points are available. 

The numerical algorithm was implemented in FORTRAN;  its output was post-processed  using Mathematica to produce graphics. 

Convergence of the numerical scheme  depends on several variables:  the numbers of spatial discretization intervals $k$, $m$ and $n$,  the size of the time step $d\tau$,   the singularity mollification parameter $\ell_0$  and the finite-interval  restriction parameters $L$ and $K$.  All of these parameters were  changed  to check for convergence  in the numerical experiments.

\subsection{Results of Numerical Integration}

The numerical integrations reported here involve the approximate solutions of $3m+3$ first-order coupled nonlinear ordinary differential equations, where $m$ is the primary spatial discretization parameter. For instance, when $m=6000$, we have a system of 18003 first-order equations that are being solved numerically. 

In the first set of numerical integrations reported here, we wish to verify numerically that the solutions of the full nonlocal and nonlinear model with $N\ne 0$ indeed approach that of the linear homogeneous $N=0$ model as the width of the initial Gaussian perturbation increases.  To this end, we let $\delta=10^{-1}$ and take $\tau_1=-\ln 2 \approx -0.7$, so that the initial epoch is taken to be $s_1=\frac{1}{2}$ corresponding to cosmological redshift  $z=1$.  We do not take $\delta$ to be very small compared to unity in this case in order to avoid numerical instabilities. The results are given in Figure 4, where the spatial maximum of $\Phi$ is plotted versus $\tau : -0.7 \to 0$.  The root-mean-square deviation from the spatial mean ($\gamma=0$) of the initial Gaussian perturbation is the \emph{effective width} of the perturbation and is given by $\Gamma \mu_0^{-1} /\sqrt{2}$. As  $\Gamma$ increases, the density of the growing mode approaches that of the linear homogeneous model in Figure 4, as theoretically predicted. 

A more ambitious numerical project would involve tackling the question of whether deviations from homogeneity of amplitude $\delta=10^{-5}$ at the initial recombination epoch with $z\approx 10^{3}$ could grow fast enough such that the maximum of the density contrast function $\Phi$ would approach unity before the present epoch thus leading to possible structure formation in cosmology. We approach this issue in Figure 5,  where the numerical results are presented in much the same way as in Figure 4.
The full nonlinear and nonlocal system~\eqref{5.3}--\eqref{5.3a} is integrated from the recombination era $\tau_1=-3 \ln {10} \approx -6.91$ till the present time $\tau=0$ with $\delta =10^{-5}$. The effective width of the initial perturbation is $\Gamma \mu_0^{-1} /\sqrt{2}$ with $\Gamma =10, 100, 1000$. As $\Gamma$ increases from the bottom graph up, the density of the growing mode approaches that of the linear homogeneous equation, which is depicted by the top graph here.  A more complete treatment of this problem is beyond the scope of our numerical work due to the limitations of our computational resources.  The numerical results presented in Figure 5 do  indicate, however, that as the effective width of the initial Gaussian perturbation increases, the density contrast may grow fast enough such that  the nonlinear perturbation could separate from the homogeneous background and collapse under its own gravity eventually to produce large scale structure in the universe.

\section{Discussion}

The persistent negative results of the experiments that have searched for the particles of dark matter~\cite{NDM1, NDM2, NDM3, NDM4} naturally lead to the notion that dark matter may not exist. That is, what appears as dark matter in astrophysics and cosmology may simply be an aspect of the gravitational interaction. In the recent classical nonlocal generalization of Einstein's theory of gravitation, it is the nonlocal feature of gravitation that simulates dark matter in astrophysics~\cite{NL1, NL2, NL3, NL4, CM, NL5, NL6, ChMa}. Dark matter is considered essential for structure formation in cosmology~\cite{Mu, GSS}; therefore, it is important to investigate whether nonlocal gravity is even capable of solving the problem of large scale cosmological structure formation.  In this paper, we take the very first step in the study of this difficult problem. No exact cosmological solution of nonlocal gravity theory is known; hence, we extend the weak-field Newtonian regime of nonlocal gravity  to the cosmological domain by assuming that the ratio of the  effective dark matter to matter decreases as the universe expands.  This supposition is consistent with the notion that nonlocality is due to the gravitational memory of past events and memory fades over time.  Within the context of such a nonlocal Newtonian cosmology, we formulate the analog of the Zel'dovich solution~\cite{Mu} of the standard Newtonian cosmology. The resulting partial integro-differential equation is analyzed and its solutions are explored. The positive results of our study indicate that the nonlocal gravity approach to the problem of structure formation in cosmology deserves further investigation.

\begin{figure}
\begin{center}
{\includegraphics[width=3in]{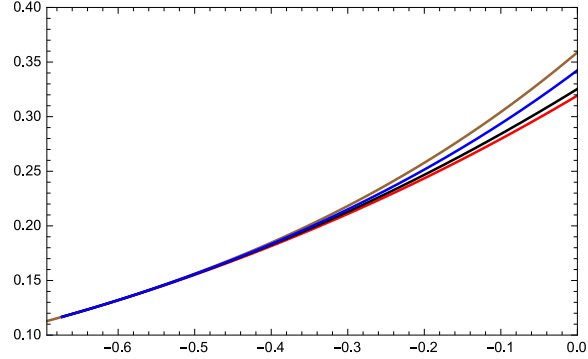}}
\end{center}
\caption{This figure depicts the numerical approximations of graphs of the maximum of  $\Phi$ as a function of $\tau$. The top graph is computed from the exact solution of the linear  homogeneous  system with $\delta=10^{-1}$ and  $\tau_1=-\ln 2 \approx -0.7$.  The lower three graphs, from the bottom up are for the nonlinear and nonlocal system with essentially the same primary initial data except that  the width of the Gaussian perturbation is scaled by $\Gamma= 10$, $\Gamma=100$ and $\Gamma=1000$, respectively. The parameters for this computation are $d\tau=0.001$, $\ell_0=10^{-6}$, $K=20.0$, $L=30, 300, 3000$ with corresponding spatial discretizations $m=1020, 3000, 6000$, respectively, and $n=24$.  The number of grid points for the integration over $\ell \in [\ell_0,K]$ is 480.  The graphs approach the exact linear  homogeneous solution for the growing mode as the width of the Gaussian perturbation  increases. \label{fig: fig4}}
\end{figure}

\begin{figure}
\begin{center}
{\includegraphics[width=3in]{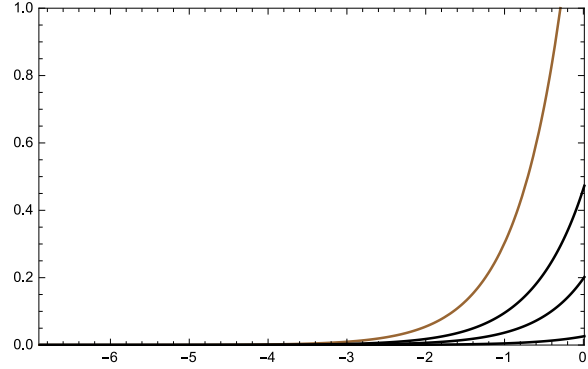}}
\end{center}
\caption{The smooth top graph is essentially the same as in Figure 3 and depicts the exact solution  of the linear homogeneous system with $\delta = 10^{-5}$ for the growing mode as a function of  $\tau: \tau_1\to 0$, where $\tau_1\approx -6.91$ at the recombination era. The three lower graphs from the bottom up depict the numerical approximations of graphs of the maximum of  $\Phi$ as functions of $\tau$ for the full nonlinear and nonlocal model with the same primary initial data except that  the width of the Gaussian perturbation is scaled by $\Gamma= 10$, $\Gamma=100$ and $\Gamma=1000$, respectively. The parameters for this computation are $d\tau=0.001$, $\ell_0=10^{-6}$, $K=20.0$, $L=30, 300, 3000$ with corresponding spatial discretizations $m = 640, 3200, 4000$, respectively, and $n=24$.  The number of grid points for the integration over $\ell \in [\ell_0,K]$ is 480.  As before, the graphs approach the exact linear  homogeneous solution for the growing mode as the width of the Gaussian perturbation  increases. \label{fig: fig5}}
\end{figure}

\begin{acknowledgments}

BM is grateful to Jeffrey Kuhn,  Roy Maartens, Sohrab Rahvar and Haojing Yan for valuable discussions.  

\end{acknowledgments}

\end{document}